%% file: Main.tex
\documentclass[journal,twoside]{IEEEtran}

\usepackage{Preamble}

\input{User}

\begin{document}


\title{%
	A Generalized Index for Static Voltage Stability
	of Unbalanced Polyphase Power Systems
	including Th{\'e}venin Equivalents and Polynomial Models%
}


\author{
	Andreas~Martin~Kettner,~\IEEEmembership{Member,~\IEEE},
	Mario~Paolone,~\IEEEmembership{Senior~Member,~\IEEE}%
	\thanks{This work was supported by the Swiss National Science Foundation (\SNSF) through the National Research Programme NRP~70 ``Energy Turnaround'', project~153997 ``Integration of Intermittent Widespread Energy Sources in Distribution Networks: Scalable and Reliable Real-Time Control of Power Flows'' (website: \protect\url{http://p3.snf.ch/project-153997}).}%
	\thanks{The authors are with the Distributed Electrical Systems Laboratory (\DESL) of the {\'E}cole Polytechnique F{\'e}d{\'e}rale de Lausanne (\EPFL), CH-1015 Lausanne, Switzerland (e-mail: \{andreas.kettner,\,mario.paolone\}@epfl.ch).}%
}

\maketitle


\markboth{Kettner and Paolone: A Generalized Index for Static Voltage Stability of Unbalanced Polyphase Power Systems}{}	




\input{Sections/Abstract}



\input{Sections/Introduction}

\input{Sections/Literature_Review}

\input{Sections/System_Model}

\input{Sections/Voltage_Stability_Index}

\input{Sections/Validation}

\input{Sections/Conclusions}



%

\bibliographystyle{IEEEtran}
\bibliography{Bibliography}



\input{Sections/Biographies}

\end{document}

%% file: User.tex
\newcommand{\IEEE}{IEEE\xspace}	
\newcommand{\SNSF}{SNSF\xspace}	
\newcommand{\EPFL}{EPFL\xspace}	
\newcommand{\DESL}{DESL\xspace}	

\newcommand{\ADN}[1][]{ADN#1\xspace}	

\newcommand{\SE}{SE\xspace}				
\newcommand{\VSA}{VSA\xspace}			
\newcommand{\CPF}{CPF\xspace}			
\newcommand{\VSI}[1][]{VSI#1\xspace}	

\newcommand{\LVR}[1][]{LVR#1\xspace}						
\newcommand{\STATCOM}[1][]{STATCOM#1\xspace}	

\newcommand{\TE}[1][]{\textup{TE#1}\xspace}		
\newcommand{\PM}[1][]{\textup{PM#1}\xspace}	
\newcommand{\CP}{CP\xspace}	
\newcommand{\CI}{CI\xspace}	
\newcommand{\CZ}{CZ\xspace}	

\newcommand{\NLP}[1][]{NLP#1\xspace}	

\newcommand{\Pie}{\Pi}
\newcommand{\Tea}{\mathrm{T}}

\newcommand{\V}{\Tensor{V}}
\newcommand{\I}{\Tensor{I}}
\newcommand{\Y}{\Tensor{Y}}
\newcommand{\Z}{\Tensor{Z}}

\newcommand{\phases}{\Set{P}}
\newcommand{\nodes}{\Set{N}}
\newcommand{\ground}{\Set{G}}
\newcommand{\branches}{\Set{L}}
\newcommand{\shunts}{\Set{T}}

\newcommand{\branchgraph}{\Graph{B}}
\newcommand{\incidence}{\Tensor{A}}

\newcommand{\Set}[1]{\mathcal{#1}}
\newcommand{\Graph}[1]{\mathscr{#1}}
\newcommand{\Cardinality}[1]{\left|#1\right|}
\newcommand{\Tensor}[1]{\boldsymbol{\mathbf{#1}}}

\newcommand{\row}{\operatorname{row}}
\newcommand{\col}{\operatorname{col}}
\newcommand{\diag}{\operatorname{diag}}

\newcommand{\Norm}[1]{\left\lVert#1\right\rVert}
\newcommand{\Kron}[1]{\widehat{#1}}

\newcommand{\diff}{\operatorname{D}}
\newcommand{\sign}{\operatorname{sign}}

{
\theoremstyle{plain}
\newtheorem{hypothesis}{Hypothesis}

\newtheorem{theorem}{Theorem}

}

{
\theoremstyle{definition}

}

%% file: Sections/Abstract.tex
\begin{abstract}
	This paper proposes a \emph{Voltage Stability Index} (\VSI) suitable for unbalanced polyphase power systems.
	To this end, the grid is represented by a polyphase multiport network model (i.e., compound hybrid parameters), and the aggregate behavior of the devices in each node by \emph{Th{\'e}venin Equivalents} (\TE[s]) and \emph{Polynomial Models} (\PM[s]), respectively.
	The proposed \VSI is a generalization of the known $L$-index, which is achieved through the use of compound electrical parameters, and the incorporation of \TE[s] and \PM[s] into its formal definition.
	Notably, the proposed \VSI can handle unbalanced polyphase power systems, explicitly accounts for voltage-dependent behavior (represented by \PM[s]), and is computationally inexpensive.
	These features are valuable for the operation of both transmission and distribution systems.
	Specifically, the ability to handle the unbalanced polyphase case is of particular value for distribution systems.
	In this context, it is proven that the compound hybrid parameters required for the calculation of the \VSI do exist under practical conditions (i.e., for lossy grids).
	The proposed \VSI is validated against state-of-the-art methods for voltage stability assessment using a benchmark system which is based on the \IEEE 34-node feeder.
\end{abstract}

%% file: Sections/Introduction.tex
\section{Introduction}
\label{Sec:Intro}


\IEEEPARstart{T}{raditional} power system control centers use elaborate tools for \emph{State Estimation} (\SE) and \emph{Voltage Stability Assessment} (\VSA) \cite{J:PSC:2005:Wu}.
As the numerical methods for solving the system equations are computationally intensive, these processes are slow.
Typically, refresh times are in the order of seconds for \SE, and minutes for \VSA.
Hence, only a few critical contingencies can be analyzed in each control cycle \cite{J:PSC:2005:Wu}.
These operational practices suffice for bulk power transmission systems, but not for power distribution systems.


Presently, the increasing penetration of distributed energy resources is pushing the development of \emph{Active Distribution Networks} (\ADN[s]).
Specifically, in order to enable the real-time operation of \ADN[s], advanced distribution management systems are required \cite{R:PSC:2011:CIGRE}.
This need has recently triggered new advances in the field.
For instance, the practical feasibility of real-time \SE has been demonstrated using phasor measurement units \cite{J:PSI:2014:Romano}, a phasor data concentrator \cite{J:PSI:2018:Derviskadic}, and a state estimator implemented in industrial hardware \cite{J:PSSE:2017:Kettner}.
The knowledge of the system state in real-time enabled the development of various real-time control methods, including hierarchical \cite{J:PSC:2011:Guerrero}, distributed \cite{J:PSC:2015:Bernstein}, and decentralized \cite{C:PSC:2016:Zhao} approaches.
To date, such methods do not perform online \VSA to guarantee static voltage stability subsequent to controller actions.
This practice is potentially dangerous.
Indeed, it has been documented that static voltage stability, rather than (as usual) thermal ratings of lines and transformers, can be the limiting operating constraint of power distribution systems (e.g., \cite{J:PSSA:PFS:1998:Prada}).
For this reason, there is a need for methods which are able to perform \VSA of power distribution systems in real-time.
Notably, in order to ensure an accurate analysis, such methods need to use realistic models of the grid (i.e., a polyphase multiport model) and the resources (i.e., incl. voltage-dependent behavior) \cite{J:PSSA:1992:Pal,J:PSSA:1993:Overbye}.


This paper proposes a \emph{Voltage Stability Index} (\VSI) based on the compound hybrid parameters of the grid, and \emph{Th{\'e}venin Equivalents} (\TE[s]) and \emph{Polynomial Models} (\PM[s]) representing the aggregate behavior of the nodes.
The proposed \VSI is a generalization of the known $L$-index \cite{J:PSSA:PFS:1986:Kessel} for more generic systems (i.e., unbalanced polyphase power systems with either radial or meshed topologies).
In this context, the contributions of this paper are threefold.
Firstly, a generalized formulation of the $L$-index, which includes a polyphase multiport network model and incorporates \TE[s] and \PM[s], is developed.
Secondly, it is proven that the required compound hybrid parameters do always exist under practical conditions (i.e., for lossy grids), thereby establishing a rigorous theoretical foundation for the $L$-index and its descendants.
Thirdly, the practical relevance of the proposed \VSI is demonstrated by validating its ability to assess the static voltage stability of a realistic power system.


The rest of this paper is organized as follows.
First, a review of the existing literature is presented in Sec~\ref{Sec:Review}.
Afterwards, the system model is described in Sec.~\ref{Sec:Model}.
The \VSI is developed in Sec.~\ref{Sec:Index}, and validated in Sec.~\ref{Sec:Validation}.
Finally, the conclusions are drawn in Sec.~\ref{Sec:Conclusion}.

%% file: Sections/Literature_Review.tex
\section{Literature Review}
\label{Sec:Review}

In order to make it easier to follow, the literature review is structured with respect to the following topics: (i) continuation power flow, (ii) maximum loadability, (iii) maximum power transfer, and (iv) power-flow solvability.
Based on this review, it is then motivated why further work is needed.


\subsection{Continuation Power Flow}
\label{Sec:Review:CPF}

Nose curves, which describe the link between active/reactive power and voltage (a.k.a. $PV$/$QV$ curves), are widely used for \VSA.
These curves are obtained via \emph{Continuation Power Flow} (\CPF) methods, which vary load or generation to produce a continuum of power-flow solutions \cite{J:PSSA:CPF:1992:Ajjarapu,J:PSSA:CPF:1995:Chiang}.
For lossy grids, static voltage instability due to generation is of little practical interest, as it occurs at excessive power injections \cite{J:PSSA:CPF:2007:Nikolaidis}.
That is, thermal ratings of lines or transformers are reached prior to instability.
In contrast, excess load can cause instability while respecting these ratings \cite{J:PSSA:PFS:1998:Prada}.
Generally, \CPF methods are computationally intensive, and too slow for real-time operation \cite{J:PSSA:CPF:1993:Canizares}, even if advanced predictors/correctors \cite{J:PSSA:CPF:2008:Li} or adaptive stepsize control \cite{J:PSSA:CPF:2012:Xu} are employed to accelerate the execution.
Usually, \CPF methods work with positive-sequence equivalent circuits of balanced three-phase systems, but the concept can be extended to unbalanced three-phase systems \cite{J:PSSA:CPF:2005:Zhang,J:PSSA:CPF:2014:Sheng}.


\subsection{Maximum Loadability}
\label{Sec:Review:ML}

Unstable operating points are bifurcations of the nonlinear system model w.r.t. nodal power absorptions or injections \cite{J:PSSA:1995:Canizares}.
These points correspond to the loadability limits of the system, which can be obtained by solving a \emph{Nonlinear Program} (\NLP), namely maximization of the loading factor (in a bus, an area, or the entire system) subject to the power-flow equations \cite{J:PSSA:ML:2009:Avalos}.
To solve this \NLP, direct or indirect iterative methods can be employed.
Direct methods explicitly consider the constraints, which means that intermediate solutions are feasible.
For example, interior-point methods \cite{J:PSSA:ML:1997:Irisarri} fall into this category.
Indirect methods instead solve a series of unconstrained optimization problems, which include penalties for constraint violations.
For example, augmented-Lagrangian methods \cite{J:PSSA:ML:1991:VanCutsem} and penalty methods \cite{C:PSSA:ML:2000:Vournas} belong to this category.


\subsection{Maximum Power Transfer}
\label{Sec:Review:MPT}

If the load is purely \emph{constant-power} (\CP), the critical point is the tip of the nose curve (i.e., the point of maximum load).
According to the maximum-power-transfer theorem, the power delivered by a source to a load reaches its maximum when the source's output impedance matches the load impedance (i.e., they are complex conjugate) \cite{J:PSSA:MPT:1973:Desoer}.
Typically, this impedance-matching criterion is applied to equivalent two-node systems, each of which consists of a single \CP load, plus a \TE of the respective external system \cite{J:PSSA:MPT:1999:Vu}.
The loads can also be represented by \PM[s], whose \emph{constant-current}~(\CI) and \emph{constant-impedance}~(\CZ) components are included into the \TE[s] \cite{J:PSSA:MPT:2003:Milosevic}.
Alternatively, the impedance-matching criterion can be used to construct the loadability surfaces of the said equivalent two-node systems
in the $PQ$ plane \cite{J:PSSA:MPT:2003:Haque,J:PSSA:MPT:2015:Vournas}.
The aforementioned approaches tacitly assume that \TE[s] reasonably reproduce the behavior of the external systems seen by the load nodes for the whole range of operating conditions, which is a priori a coarse approximation.
Hence, some researchers advocate the use of more elaborate models, which are based on so-called coupled single-port circuits.
For example, the \TE can be extended by a coupling term \cite{J:PSSA:MPT:2011:Wang} or refined with sensitivity coefficients \cite{J:PSSA:MPT:2017:Cui}.
Finally, Ward equivalents can be used instead of \TE[s] \cite{J:PSSA:MPT:2014:Liu}.


\subsection{Power-Flow Solvability}
\label{Sec:Review:PFS}

The power-flow equations are borderline (un)solvable on the loadability surface.
It is possible to formulate conditions for the solvability of the power-flow equations \cite{J:PSSA:PFS:2012:Grijalva}, or to construct approximations of the loadability surface \cite{J:PSSA:PFS:2016:Bolognani,J:PSSA:PFS:2017:Machado,J:PSSA:PFS:2018:Wang}.
However, these approaches tend to be computationally intricate.
Hence, most works instead exploit that Jacobian matrix of the power-flow equations is singular on the loadability surface \cite{J:PSSA:PFS:1990:Sauer}.
More precisely, the determinant \cite{J:PSSA:PFS:1998:Prada}, eigenvalues \cite{J:PSSA:PFS:1992:Gao}, and singular values \cite{J:PSSA:PFS:1993:Loef} of the Jacobian are widely used as \VSI[s].
Another popular family of \VSI[s] descends from the $L$-index \cite{J:PSSA:PFS:1986:Kessel}, which is derived from the hybrid parameters of the grid.
The original $L$-index \cite{J:PSSA:PFS:1986:Kessel} is based on idealized models of generators (i.e., constant-voltage sources) and loads (i.e., \CP loads), but there exist variants based on more generic models of generators (i.e., \TE[s]) \cite{J:PSSA:PFS:2013:Wang} or loads (i.e., \PM[s]) \cite{J:PSSA:PFS:2005:Hongjie}.
Lastly, note that most \VSI[s] vary nonlinearly with the load.
That is, there may not be a one-to-one relation between \VSI and loadability margin.
However, for special cases, such as \CP loads \cite{J:PSSA:PFS:1997:ElKateb} or \PM[s] with constant power factor \cite{C:PSSA:PFS:2015:Matavalam}, \VSI[s] with more linear behavior do exist.


\subsection{Motivation for Further Work}

The evolution towards \ADN[s] has sparked the development of various methods for real-time control (e.g., \cite{J:PSC:2011:Guerrero,J:PSC:2015:Bernstein,C:PSC:2016:Zhao}).
Yet, to date, such control methods do not perform online \VSA to ensure stable operation subsequent to a control decision.
This negligence is dangerous, since static voltage instability is a proven threat in distribution systems (e.g., \cite{J:PSSA:PFS:1998:Prada}).
Moreover, the unbalanced polyphase nature of the grid is normally ignored.
Therefore, there is a need for \VSA tools which are capable of real-time operation (unlike Sec. \ref{Sec:Review:CPF}/\ref{Sec:Review:ML}), and able to handle a detailed system model (unlike Sec. \ref{Sec:Review:MPT}/\ref{Sec:Review:PFS}).
To this end, this paper proposes a generalized formulation of the $L$-index, which is based on the compound hybrid parameters of the grid, and \TE[s] and \PM[s] of the nodes.
These equivalents are suitable for representing diverse distributed loads and generators, with or without power electronic interfaces \cite{J:PSMI:1988:Price,J:PSMI:1998:Hajagos}.

%% file: Sections/System_Model.tex
\section{System Model}
\label{Sec:Model}


\subsection{Electrical Grid}
\label{Sec:Model:Grid}

Subsequently, the grid model developed in \cite{J:PSM:2018:Kettner} is recalled.


Consider an unbalanced polyphase power system equipped with a neutral conductor.
The system is wired as follows:
\begin{hypothesis}
	\label{Hyp:Neutral}
	The neutral conductor is grounded through an effective earthing system, which establishes a null voltage w.r.t. the ground.
	Moreover, the reference points of all sources (i.e., voltage or current) are connected to the neutral conductor.
\end{hypothesis}
Under these conditions, the phase-to-neutral voltages are de facto phase-to-ground voltages, and fully describe the system.
The phases are numbered as $p\in\phases\coloneq\{1,\cdots,\Cardinality{\phases}\}$, and the ground node as $g\in\ground\coloneq\{0\}$.
A \emph{polyphase node} is a complete set of phase terminals that belong together.
The clamps of the electrical components which the grid is built from (e.g., lines and transformers) form the set of \emph{physical} polyphase nodes $\nodes_{\text{physical}}$.
As to the grid, the following hypothesis is made
\begin{hypothesis}
	\label{Hyp:Components}
	The grid consists of linear passive components.
	In a per-unit model, these components can be represented by polyphase $\Pie$-section or $\Tea$-section two-port equivalent circuits, whose branch and shunt elements are described by compound impedance and admittance matrices, respectively.
\end{hypothesis}
\noindent
That is, only the electromagnetic coupling within components, but not between them, is considered.

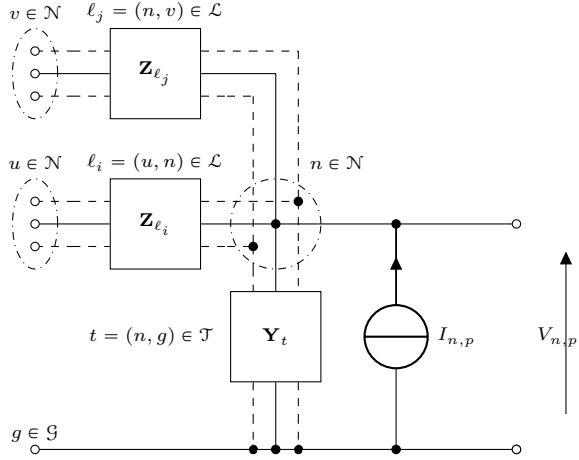
\begin{figure}[t]
	\centering
	\input{Figures/Model_Grid}
	\caption{Definition of the compound branch impedance matrices $\Z_{\ell}$ ($\ell\in\branches$), compound shunt admittance matrices $\Y_{t}$ ($t\in\shunts$), nodal voltage phasors $V_{n,p}$ ($n\in\nodes$, $p\in\phases$), and injected current phasors $I_{n,p}$ ($n\in\nodes$, $p\in\phases$).}
	\label{Fig:Model:Grid}
\end{figure}

These equivalent circuits may introduce \emph{virtual} polyphase nodes $\nodes_{\text{virtual}}$.
For instance, every $\Tea$-section equivalent circuit adds one virtual polyphase node.
Let $\nodes\coloneq\nodes_{\text{physical}}\cup\nodes_{\text{virtual}}$ be the set of all polyphase nodes.
The topology of the grid model is described by the \emph{polyphase branches} $\ell\in\branches\subseteq\nodes\times\nodes$ and the \emph{polyphase shunts} $t\in\shunts\coloneq\nodes\times\ground$.
The \emph{branch graph} $\branchgraph\coloneq(\nodes,\branches)$ is described by the \emph{branch incidence matrix} $\incidence_{\branchgraph}$
\begin{equation}
	\incidence_{\branchgraph}:~A_{\branchgraph,kn}
	\coloneq
		\left\{
		\begin{array}{rc}
			+1	&\text{if}~\ell_{k}=(n,\makebox[1ex]{$\cdot$})\in\branches\\
			-1	&\text{if}~\ell_{k}=(\makebox[1ex]{$\cdot$},n)\in\branches\\
			0	&\text{otherwise}
		\end{array}
		\right.
\end{equation}
Note that $\incidence_{\branchgraph}$ exists for any topology (i.e., radial and meshed).
Every polyphase branch $\ell\in\branches$ is associated with a \emph{compound branch impedance matrix} $\Z_{\ell}$, and every polyphase shunt $t\in\shunts$ with a \emph{compound branch admittance matrix} $\Y_{t}$ (see Fig.~\ref{Fig:Model:Grid}).
Regarding these parameters, the following hypothesis is made
\begin{hypothesis}
	\label{Hyp:Parameters}
	For all polyphase branches $\ell\in\branches$, it holds that
	\begin{equation}
		\Z_{\ell}=\Z_{\ell}^{T},~
		\Re\{\Z_{\ell}\}\succeq0,~
		\exists\Y_{\ell}\coloneq\Z_{\ell}^{-1}
		\label{Eq:Parameters:Branch}
	\end{equation}
	For all polyphase shunts $t\in\shunts$ with $\Y_{t}\neq\Tensor{0}$, it holds that
	\begin{equation}
		\Y_{t}=\Y_{t}^{T},~
		\Re\{\Y_{t}\}\succeq0,~
		\exists\Z_{t}\coloneq\Y_{t}^{-1}
		\label{Eq:Parameters:Shunt}
	\end{equation}%
\end{hypothesis}
\noindent%
Note that $\Re\{\Z_{\ell}\}\succeq0$ and $\Re\{\Y_{t}\}\succeq0$ imply lossiness.

Let $V_{n,p}$ and $I_{n,p}$ denote the phasors of the phase-to-ground voltage and injected current in phase $p$ of node $n$, respectively (see Fig.~\ref{Fig:Model:Grid}).
Define
\begin{alignat}{2}
	\V			&\coloneqq\col_{n\in\nodes}(\V_{n})	,	&~
	\V_{n}	&\coloneq\col_{p\in\phases}(V_{n,p})
	\\
	\I			&\coloneqq\col_{n\in\nodes}(\I_{n}),		&~
	\I_{n}		&\coloneq\col_{p\in\phases}(I_{n,p})
\end{alignat}
The \emph{compound admittance matrix} $\Y$ describes Ohm's law
\begin{equation}
	\I=\Y\V
	\label{Eq:Ohm}
\end{equation}
Define the \emph{polyphase incidence matrix} $\incidence^{\phases}_{\branchgraph}$ and the \emph{primitive compound admittance matrices} $\Y_{\branches}$ and $\Y_{\shunts}$ as
\begin{align}
	\incidence^{\phases}_{\branchgraph}
	&\coloneq	\incidence_{\branchgraph}\otimes\diag(\Tensor{1}_{\Cardinality{\phases}\times1})
	\\
	\Y_{\branches}
	&\coloneq	\diag_{\ell\in\branches}(\Y_{\ell})
	\\
	\Y_{\shunts}
	&\coloneq	\diag_{t\in\shunts}(\Y_{t})
\end{align}
where $\Tensor{1}_{M\times N}$ is a matrix of ones with size $M\times N$, and $\otimes$ is the \emph{Kronecker product}.
Then, $\Y$ is constructed as follows
\begin{equation}
	\Y
	=	(\incidence^{\phases}_{\branchgraph})^{T}\Y_{\branches}\incidence^{\phases}_{\branchgraph}+\Y_{\shunts}
\end{equation}
Let $\Set{A},\Set{B}\subsetneq\nodes$ so that $\Set{A}\cap\Set{B}=\emptyset$.
Define $\I_{\Set{A}} \coloneq \col_{n\in\Set{A}}(\I_{n})$, $\V_{\Set{B}} \coloneq \col_{n\in\Set{B}}(\V_{n})$, and $\Y_{\Set{A}\times\Set{B}}$ as the block of $\Y$ that relates $\I_{\Set{A}}$ and $\V_{\Set{B}}$.
The following properties hold (see \cite{J:PSM:2018:Kettner} for proof).
\begin{theorem}
	\label{Thm:Kron}
	Let $\Set{Z}\subsetneq\nodes$, s.t. $\Set{Z}\neq\emptyset$ and $\I_{\Set{Z}}=\Tensor{0}$ (i.e., $\Set{Z}$ has zero injected currents).
	Define $\Set{Z}_{\complement}\coloneq\nodes\setminus\Set{Z}$.
	If Hypotheses \ref{Hyp:Neutral}--\ref{Hyp:Parameters} hold, $\branchgraph$ is weakly connected, and $\Re\{\Z_{\ell}\}\succ0$ $\forall\ell\in\branches$, then Ohm's law \eqref{Eq:Ohm} can be reduced to the following form
	\begin{equation}
		\I_{\Set{Z}_{\complement}} 
		=	\widehat{\Y}\V_{\Set{Z}_{\complement}}
		,~
		\Kron{\Y} = \Y / \Y_{\Set{Z}\times\Set{Z}}
		\label{Eq:Kron}
	\end{equation}
	where $\Y / \Y_{\Set{Z}\times\Set{Z}}$ is the \emph{Schur complement} of $\Y$ w.r.t. $\Y_{\Set{Z}\times\Set{Z}}$.
	If $\Set{Z}$ is partitioned as $\{\Set{Z}_{k}\,|\,k\in\Set{K}\}$, the $\Set{Z}_{k}$ can also be reduced one after another (i.e., in sequence rather than in parallel).
\end{theorem}
\begin{theorem}
	\label{Thm:Hybrid}
	Let $\Set{M}\subsetneq\nodes$ s.t. $\Set{M}\neq\emptyset$.
	If Hypotheses \ref{Hyp:Neutral}--\ref{Hyp:Parameters} hold, $\branchgraph$ is weakly connected, and $\Re\{\Z_{\ell}\}\succ0$ $\forall\ell\in\branches$, then there exists a \emph{compound hybrid matrix} $\Tensor{H}$ so that
	\begin{equation}
		\left[
		\begin{array}{c}
			\I_{\Set{M}_{\complement}}\\
			\V_{\Set{M}}
		\end{array}
		\right]
		=	\left[
			\begin{array}{ll}
					\Tensor{H}_{\Set{M}_{\complement}\times\Set{M}_{\complement}}
				&	\Tensor{H}_{\Set{M}_{\complement}\times\Set{M}}
				\\
					\Tensor{H}_{\Set{M}\times\Set{M}_{\complement}}
				&	\Tensor{H}_{\Set{M}\times\Set{M}}
			\end{array}
			\right]
			\left[
			\begin{array}{c}
				\V_{\Set{M}_{\complement}}\\
				\I_{\Set{M}}
			\end{array}
			\right]
			\label{Eq:Hybrid}
	\end{equation}
	whose blocks are given by
	\begin{alignat}{2}
		&		\Tensor{H}_{\Set{M}\times\Set{M}}
		&&=	\phantom{-}\Y_{\Set{M}\times\Set{M}}^{-1}
		\\
		&		\Tensor{H}_{\Set{M}\times\Set{M}_{\complement}}
		&&=	-\Y_{\Set{M}\times\Set{M}}^{-1}
				\Y_{\Set{M}\times\Set{M}_{\complement}}
		\\
		&		\Tensor{H}_{\Set{M}_{\complement}\times\Set{M}}
		&&=	\phantom{-}\Y_{\Set{M}_{\complement}\times\Set{M}}
				\Y_{\Set{M}\times\Set{M}}^{-1}
		\\
		&		\Tensor{H}_{\Set{M}_{\complement}\times\Set{M}_{\complement}}
		&&=	\phantom{-}\Y / \Y_{\Set{M}\times\Set{M}}
	\end{alignat}
	This property holds both for unreduced and (partially) reduced compound admittance matrices (i.e., $\Y$ and $\Kron{\Y}$ in Theorem~\ref{Thm:Kron}).%
\end{theorem}


\subsection{Aggregate Behavior of the Nodes}
\label{Sec:Model:Node}

\begin{figure}[t]
	\centering
	
	\subfloat[]
	{%
		\centering
		\input{Figures/Model_Node_Thevenin}
		\label{Fig:Node:Thevenin}
	}
	$~$
	\subfloat[]
	{%
		\centering
		\input{Figures/Model_Node_Polynomial}
		\label{Fig:Node:Polynomial}
	}
	
	\caption{Representation of the aggregate node behaviour: (\ref{Fig:Node:Thevenin}) \TE of a slack node $s\in\Set{S}$, (\ref{Fig:Node:Polynomial}) \PM of phase $p\in\phases$ in a resource node $r\in\Set{R}$.}
	\label{Fig:Node}
\end{figure}
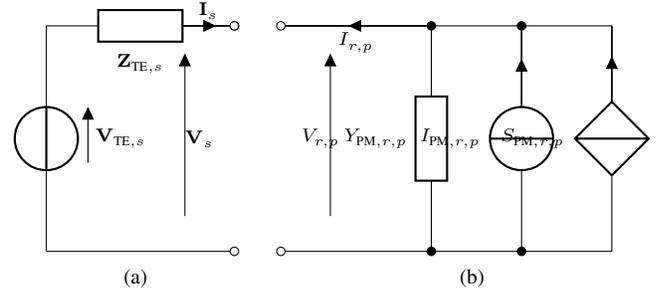

The nodes are divided into three sets based on their generic behaviour.
Namely, $\nodes=\Set{Z}\cup\Set{S}\cup\Set{R}$, where $\Set{Z}$ stands for \emph{zero-injection} nodes,
$\Set{S}$ for \emph{slack} nodes, and $\Set{R}$ for \emph{resource} nodes.


In zero-injection nodes, there are no devices.
Hence
\begin{equation}
	\I_{\Set{Z}} = \Tensor{0}
	\label{Eq:Node:Zero}
\end{equation}


At the slack nodes, the voltage (and frequency) is regulated, either by a device, for instance a synchronous machine \cite{J:PSSA:PFS:2013:Wang} or a power electronic device \cite{J:PSC:2012:Rocabert}, or a link to the main grid.
Accordingly, the slack nodes $s\in\Set{S}$ behave as non-ideal voltage sources, which can be represented by \TE[s] \cite{J:PSSA:PFS:2013:Wang}:
\begin{equation}
	\V_{s} = \V_{\TE,s} - \Z_{\TE,s}\I_{s}
	\label{Eq:Thevenin}
\end{equation}
where $\V_{\TE,s}$ and $\Z_{\TE,s}$ are the \TE voltages and impedances, respectively (see Fig.~\ref{Fig:Node:Thevenin}).


At the resource nodes, non-zero power is injected/absorbed, but the voltage is not regulated.
This behaviour corresponds to voltage-dependent power sources, which can be approximated by \PM[s] \cite{J:PSMI:1988:Price,J:PSMI:1998:Hajagos}.
Define the normalized voltage $v_{r,p}$ in phase $p\in\Set{R}$ of resource node $r\in\Set{R}$ as
\begin{equation}
	v_{r,p} \coloneq \frac{V_{r,p}}{V_{0,r}}
\end{equation}
where $V_{0,r}$ is a given reference voltage (e.g., nominal voltage).
Assuming that the equivalent power sources have no coupling among the phases, the injected active powers $P_{r,p}$ and reactive powers $Q_{r,p}$ are given by quadratic polynomials of the $v_{r,p}$:
\begin{align}
	P_{r,p}	&=	\lambda_{r,p}P_{0,r,p}(\alpha_{\Re,r,p}|v_{r,p}|^{2}+\beta_{\Re,r,p}|v_{r,p}|+\gamma_{\Re,r,p})
	\label{Eq:Polynomial:P}
	\\
	Q_{r,p}	&=	\lambda_{r,p}Q_{0,r,p}(\alpha_{\Im,r,p}|v_{r,p}|^{2}+\beta_{\Im,r,p}|v_{r,p}|+\gamma_{\Im,r,p})
	\label{Eq:Polynomial:Q}
\end{align}
where $\alpha_{\Re/\Im}$, $\beta_{\Re/\Im}$, and $\gamma_{\Re/\Im}$ are normalized coefficients (i.e., $\alpha_{\Re/\Im}+\beta_{\Re/\Im}+\gamma_{\Re/\Im}=1$), $\lambda$ is a loading factor, and $P_{0}$ and $Q_{0}$ are reference powers which correspond to $\lambda=1$ and $|v|=1$.
In general, as indicated by the subscripts $r$ and $p$ in \eqref{Eq:Polynomial:P}--\eqref{Eq:Polynomial:Q}, the aforestated quantities are functions of the node and phase.
For given $\lambda_{r,p}$, $S_{r,p}=P_{r,p}+jQ_{r,p}$ can be written as
\begin{equation}
	S_{r,p} \approx -Y_{\PM,r,p}^{*}|V_{r,p}|^{2} + V_{r,p}I_{\PM,r,p}^{*} + S_{\PM,r,p}
	\label{Eq:Polynomial:S}
\end{equation}
where $Y_{\PM,r,p}$, $I_{\PM,r,p,}$, and $S_{\PM,r,p}$ are \CZ, \CI, and \CP terms, respectively (see Fig.~\ref{Fig:Node:Polynomial}).
Recall from Sec.~\ref{Sec:Review:MPT} that, provided that the load is purely \CP, the critical point lies at the tip of the nose curve.
If the load contains \CI or \CZ components, this is not the case.
Namely, injection or absorption terms shift the critical point to the upper or lower portion of the nose curve, respectively \cite{J:PSSA:1993:Overbye}.
	

The parameters of the \TE[s] and \PM[s] can be derived formally or numericall, if white-box models of the underlying devices are available.
In practice, it is often easier to estimate them from measurements, for instance using weighted-least-squares regression \cite{J:PSMI:2013:Dzafic}.
In this paper, it is assumed that the model parameters are known -- irrespective of how they are obtained.

%% file: Figures/Model_Grid.tex
\tikzstyle{block}=[rectangle, draw=black, minimum size=1.2cm, inner sep=0.0cm]
\tikzstyle{elliptic}=[ellipse, dashdotted, draw=black, minimum width=0.6cm, minimum height=1.2cm, inner sep=0.0cm]
\tikzstyle{circular}=[circle, dashdotted, draw=black, minimum size=1.2cm, inner sep=0.0cm]

\begin{circuitikz}
	\scriptsize
	
	\def\BranchPosition{1.6}	
	\def\BranchHeightLower{3.0}	
	\def\BranchHeightUpper{5.0}	
	\def\ShuntHeight{1.5} 
	\def\PortPosition{3.2}	
	\def\BlockSize{1.2}	
	\def\WireSpacing{0.3}	
	
	
	\coordinate (GL) at (-\PortPosition,0);	
	\coordinate (GC) at (0,0);					
	\coordinate (GR) at (\PortPosition,0);	
	
	\draw (GL)
		to[short,o-] (GC)
		to[short,-o] (GR);
	
	
	\node[elliptic] (TLB) at (-\PortPosition,\BranchHeightLower) {};
	\node[circular] (ICB) at (0,\BranchHeightLower) {};	
	\node[block] (ZB) at (-\BranchPosition,\BranchHeightLower) {$\Z_{\ell_{i}}$};	
	\node[block] (Y) at (0,\ShuntHeight) {$\Y_{t}$};
		
	\draw (TLB)
		to[short,o-] (ZB.west)
		to[open] (ZB.east)
		to[short] (ICB)
		to[short] (Y.north)
		to[open] (Y.south)
		to[short,-*] (GC);
	\draw[dashed] ($(TLB)+\WireSpacing*(0,1)$)
		to[short] ($(ZB.west)+\WireSpacing*(0,1)$)
		to[open] ($(ZB.east)+\WireSpacing*(0,1)$)
		to[short] ($(ICB)+\WireSpacing*(1,1)$)
		to[short] ($(Y.north)+\WireSpacing*(1,0)$)
		to[open] ($(Y.south)+\WireSpacing*(1,0)$)
		to[short,-*] ($(GC)+\WireSpacing*(1,0)$);
	\draw[dashed] ($(TLB)-\WireSpacing*(0,1)$)
		to[short] ($(ZB.west)-\WireSpacing*(0,1)$)
		to[open] ($(ZB.east)-\WireSpacing*(0,1)$)
		to[short] ($(ICB)-\WireSpacing*(1,1)$)
		to[short] ($(Y.north)-\WireSpacing*(1,0)$)
		to[open] ($(Y.south)-\WireSpacing*(1,0)$)
		to[short,-*] ($(GC)-\WireSpacing*(1,0)$);
	\draw($(TLB)+\WireSpacing*(0,1)$) to[open,o-o] ($(TLB)-\WireSpacing*(0,1)$);
	
	\node at ($(TLB.north)+\WireSpacing*(0,0.7)$) {$u\in\nodes$};
	\node at ($(ICB.north)+\WireSpacing*(2.7,0.7)$) {$n\in\nodes$};
	\node at ($(GL)+\WireSpacing*(0,0.7)$) {$g\in\ground$};
	\node at ($(ZB.north)+\WireSpacing*(0,0.7)$) {$\ell_{i}=(u,n)\in\branches$};
	\node at ($(Y.west)-\WireSpacing*(3.5,0)$) {$t=(n,g)\in\shunts$};
	
	
	\node[elliptic] (TLA) at (-\PortPosition,\BranchHeightUpper) {};
	\node[block] (ZA) at (-\BranchPosition,\BranchHeightUpper) {$\Z_{\ell_{j}}$};
	\coordinate (ICA) at (0,\BranchHeightUpper);	
	
	\draw (TLA)
		to[short,o-] (ZA.west)
		to[open] (ZA.east)
		to[short] (ICA)
		to[short,-*] (ICB);
	\draw[dashed] ($(TLA)+\WireSpacing*(0,1)$)
		to[short] ($(ZA.west)+\WireSpacing*(0,1)$)
		to[open] ($(ZA.east)+\WireSpacing*(0,1)$)
		to[short] ($(ICA)+\WireSpacing*(1,1)$)
		to[short] ($(ICB)+\WireSpacing*(1,1)$);
	\draw[dashed] ($(TLA)-\WireSpacing*(0,1)$)
		to[short] ($(ZA.west)-\WireSpacing*(0,1)$)
		to[open] ($(ZA.east)-\WireSpacing*(0,1)$)
		to[short] ($(ICA)-\WireSpacing*(1,1)$)
		to[short] ($(ICB)-\WireSpacing*(1,1)$);
	\draw($(TLA)+\WireSpacing*(0,1)$) to[open,o-o] ($(TLA)-\WireSpacing*(0,1)$);
	\draw($(ICB)+\WireSpacing*(1,1)$) to[open,*-*] ($(ICB)-\WireSpacing*(1,1)$);
	
	\node at ($(TLA.north)+\WireSpacing*(0,0.7)$) {$v\in\nodes$};
	\node at ($(ZA.north)+\WireSpacing*(0,0.7)$) {$\ell_{j}=(n,v)\in\branches$};
	
	
	\coordinate (TR) at (\PortPosition,\BranchHeightLower);
	
	\draw (TR) to[short] (ICB);
	
	\draw (GR) to[open,v=$V_{n,p}$,o-o] (TR);
	\draw ($(0.5*\PortPosition,0)$) to[current source,i_=$I_{n,p}$,*-*] ($(0.5*\PortPosition,\BranchHeightLower)$);
	
\end{circuitikz}

%% file: Figures/Model_Node_Thevenin.tex
\begin{circuitikz}
	\scriptsize
	
	\def\xext{0.0}
	\def\xint{2.5};
	\def\y{3.0};
	
	\coordinate (TIU) at (-\xint,\y);
	\coordinate (TIL) at (-\xint,0);
	
	\coordinate (TEU) at (-\xext,\y);
	\coordinate (TEL) at (-\xext,0);
	
	\draw (TIL) to[voltage source,v_=$\V_{\TE,s}$] (TIU);
	\draw (TIU) to[generic,i=$\I_{s}$,-o] (TEU);
	\node at ($0.5*(TIU)+0.5*(TEU)-(0,0.5)$) {$\Z_{\TE,s}$}; 
	\draw (TEL) to[short] (TIL);
	\draw (TEL) to[open,o-o,v^=$\V_{s}$] (TEU);
	
\end{circuitikz}

%% file: Figures/Model_Node_Polynomial.tex
\begin{circuitikz}
	\scriptsize
	
	\def\xeq{3.2};
	\def\y{3.0};
	
	\coordinate (NRU) at (0,\y);
	\coordinate (NRL) at (0,0);

	\coordinate (CIU) at (\xeq,\y);
	\coordinate (CIL) at (\xeq,0);
		
	\coordinate (CYU) at ($(CIU)-(1.2,0)$);
	\coordinate (CYL) at ($(CIL)-(1.2,0)$);
	
	\coordinate (CPU) at ($(CIU)+(1.2,0)$);
	\coordinate (CPL) at ($(CIL)+(1.2,0)$);
	
	\draw (NRL) to[open,v=$V_{r,p}$] (NRU);
	
	\draw (NRL) to[short,o-] (CYL) to[generic=$Y_{\PM,r,p}$] (CYU) to[short,-o,i=$I_{r,p}$] (NRU);
	\draw (CYL) to[short,*-] (CIL) to[current source,i=$I_{\PM,r,p}$] (CIU) to[short,-*] (CYU);
	\draw (CIL) to[short,*-] (CPL) to[controlled current source,i=$S_{\PM,r,p}$] (CPU) to[short,-*] (CIU);
	
\end{circuitikz}

%% file: Sections/Voltage_Stability_Index.tex
\section{Voltage Stability Index}
\label{Sec:Index}

In the following, the generalized $L$-index is developed based on the aforementioned models.
To this end, a procedure similar to the derivation of the original formulation of the $L$-index \cite{J:PSSA:PFS:1986:Kessel} is followed.
Namely, the equations describing the polyphase network \eqref{Eq:Ohm}, the \TE[s] \eqref{Eq:Thevenin}, and the \PM[s] \eqref{Eq:Polynomial:S} are combined to yield a complex quadratic equation.


\begin{figure}
	\centering
	\input{Figures/Model_System}
	\caption{Schematic of the system model with augmented electrical grid.}
	\label{Fig:Model:System}
\end{figure}
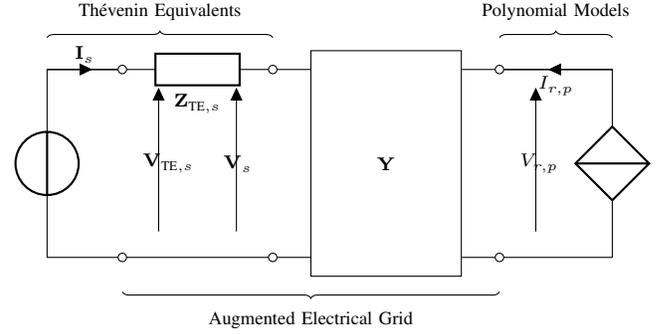

Summarizing Sec.~\ref{Sec:Model}, the system is described by (see Fig.~\ref{Fig:Model:System})
\begin{align}
	\I
	&=	\Y\V
	\label{Eq:System:Original:Ohm}
	\\
	\I_{\Set{Z}}
	&=	\Tensor{0}
	\label{Eq:System:Original:Zero}
	\\
	\V_{s}
	&=	\V_{\TE,s} - \Z_{\TE,s}\I_{s},~
		\forall s\in\Set{S}
	\label{Eq:System:Original:Thevenin}
	\\
	I_{r,p}
	&=	-Y_{\PM,r,p}V_{r,p} + I_{\PM,r,p} + \frac{S_{\PM,r,p}^{*}}{V_{r,p}^{*}},~
		\begin{array}{l}
			\forall r\in\Set{R},\\
			\forall p\in\Set{P}
		\end{array}
	\label{Eq:System:Original:Polynomial}
\end{align}
If it holds that
\begin{hypothesis}
	\label{Hyp:Thevenin}
	The impedances $\Z_{\TE,s}$ ($s\in\Set{S}$) satisfy
	\begin{equation}
		\Z_{\TE,s}=\Z_{\TE,s}^{T},~
		\Re\{\Z_{\TE,s}\}\succeq0,~
		\exists\Y_{\TE,s}\coloneq\Z_{\TE,s}^{-1}
	\end{equation}
\end{hypothesis}
\noindent
which is the analogon of \eqref{Eq:Parameters:Branch}, the model can be reinterpreted.
Define $\Set{I}$ as the set of \emph{internal} nodes of the \TE[s].
The equivalent voltage sources of the \TE[s] and the \PM[s] are connected through the \emph{augmented} electrical grid with nodes $\Set{N}'\coloneq\Set{I}\cup\Set{N}$, which is composed of the physical electrical grid and the equivalent impedances of the \TE[s] (see Fig.~\ref{Fig:Model:System}).
The augmented electrical grid can also be described by Ohm's law, namely
\begin{equation}
	\I'=\Y'\V'
	\label{Eq:System:Augmented:Ohm}
\end{equation}
where $\I'$ and $\V'$ are the vectors of injected currents and phase-to-ground voltages, and $\Y'$ is the compound admittance matrix of the augmented electrical grid.
More precisely, \eqref{Eq:System:Augmented:Ohm} is obtained by combining \eqref{Eq:System:Original:Ohm} with \eqref{Eq:System:Original:Thevenin}.
Define
\begin{align}
	\V_{\TE}
	&\coloneq	\col_{s\in\Set{S}}(\V_{\TE,s})
	\\
	\Y_{\TE}
	&\coloneq 	\diag_{s\in\Set{S}}(\Y_{\TE,s})
\end{align}
where Hypothesis~\ref{Hyp:Thevenin} ensures the existence of the $\Y_{\TE,s}$ $\forall s\in\Set{S}$.
Furthermore, let $\V'$ and $\I'$ be constructed as follows\footnote{The elements of $\V'$ and $\I'$ can be ordered arbitrarily. This particular order is convenient, because it results in a well-arranged $\Y'$.}
\begin{alignat}{2}
	\V'
	&=	\left[
			\begin{array}{l}
				\V'_{\Set{I}}\\
				\V'_{\Set{S}}\\
				\V'_{\Set{Z}}\\
				\V'_{\Set{R}}
			\end{array}
			\right]
	&&=	\left[
			\begin{array}{l}
				\V_{\text{\TE}}\\
				\V_{\Set{S}}\\
				\V_{\Set{Z}}\\
				\V_{\Set{R}}
			\end{array}
			\right]
	\\
	\I'
	&=	\left[
			\begin{array}{l}
				\I'_{\Set{I}}\\
				\I'_{\Set{S}}\\
				\I'_{\Set{Z}}\\
				\I'_{\Set{R}}
			\end{array}
			\right]
	&&=	\left[
			\begin{array}{l}
				\I_{\Set{S}}\\
				\Tensor{0}\\
				\Tensor{0}\\
				\I_{\Set{R}}
			\end{array}
			\right]
\end{alignat}
Observe that the slack nodes $\Set{S}$ are zero-injection nodes in the augmented electrical grid (i.e., $\I'_{\Set{S}}=\Tensor{0}$).
Through combination of \eqref{Eq:System:Original:Ohm} and \eqref{Eq:System:Original:Thevenin}, $\Y'$ is obtained as
\begin{equation}
	\Y'
	=	\left[
		\begin{array}{cccc}
				\phantom{-}\Y_{\TE}
			&	\multicolumn{1}{l}{-\Y_{\TE}}
			&	\Tensor{0}
			&	\Tensor{0}
			\\
				-\Y_{\TE}
			&	\multicolumn{1}{r}{\hphantom{-}\Y_{\TE}+\Y_{\Set{S}\times\Set{S}}}
			&	\Y_{\Set{S}\times\Set{Z}}
			&	\Y_{\Set{S}\times\Set{R}}
			\\
				\Tensor{0}
			&	\multicolumn{1}{r}{\Y_{\Set{S}\times\Set{Z}}}
			&	\Y_{\Set{Z}\times\Set{Z}}
			&	\Y_{\Set{Z}\times\Set{R}}
			\\
				\Tensor{0}
			&	\multicolumn{1}{r}{\Y_{\Set{S}\times\Set{R}}}
			&	\Y_{\Set{R}\times\Set{Z}}
			&	\Y_{\Set{R}\times\Set{R}}
			\\
		\end{array}	
		\right]
\end{equation}


If Hypotheses~\ref{Hyp:Neutral}--\ref{Hyp:Thevenin} hold, then the augmented electrical grid satisfies the conditions of Theorems~\ref{Thm:Kron}--\ref{Thm:Hybrid}.
Thus, the nodes $\Set{S}\cup\Set{Z}$ can be eliminated via Kron reduction, which yields a reduced electrical grid, which is described by (see Theorem~\ref{Thm:Kron})
\begin{equation}
	\left[
	\begin{array}{l}
		\I_{\Set{S}}\\
		\I_{\Set{R}}
	\end{array}
	\right]
	=	\Kron{\Y}'
		\left[
		\begin{array}{l}
			\V_{\TE}\\
			\V_{\Set{R}}
		\end{array}
		\right]
	,~
	\Kron{\Y}'
	=	\Y' / \Y'_{\{\Set{S}\cup\Set{Z}\}\times\{\Set{S}\cup\Set{Z}\}}
\end{equation}
The above equation can be reformulated as (see Theorem~\ref{Thm:Hybrid})
\begin{equation}
	\left[
	\begin{array}{c}
		\I_{\Set{S}}	\\
		\V_{\Set{R}}
	\end{array}
	\right]
	=	\left[
		\begin{array}{ll}
				\Kron{\Tensor{H}}'_{\Set{I}\times\Set{I}}
			&	\Kron{\Tensor{H}}'_{\Set{I}\times\Set{R}}
			\\
				\Kron{\Tensor{H}}'_{\Set{R}\times\Set{I}}
			&	\Kron{\Tensor{H}}'_{\Set{R}\times\Set{R}}
		\end{array}
		\right]
		\left[
		\begin{array}{c}
			\V_{\TE}\\
			\I_{\Set{R}}
		\end{array}
		\right]
	\label{Eq:System:Augmented:Hybrid}
\end{equation}
From the second block row, it follows that
\begin{align}
	V_{r,p}
	&=		\widetilde{V}_{\TE,r,p}
			+	\sum\limits_{j\in\Set{R}}\row_{p}(\Kron{\Tensor{H}}'_{\mathit{rj}})\I_{j},~
			\begin{array}{l}
				\forall r\in\Set{R},\\
				\forall p\in\Set{P}
			\end{array}
	\label{Eq:V:1}
	\\
	\widetilde{V}_{\TE,r,p}
	&\coloneq	\sum\limits_{i\in\Set{S}}\row_{p}(\Kron{\Tensor{H}}'_{\mathit{ri}})\V_{\TE,i}
\end{align}
Recall that the elements $I_{j,q}$ of $\I_{j}$ ($j\in\Set{R}$, $q\in\Set{P}$) are given by \eqref{Eq:System:Original:Polynomial}.
Express $I_{j,q}$ explicitly as a function of $V_{r,p}$.
Namely
\begin{align}
	I_{j,q}
	&=	-\widetilde{Y}_{\PM,j,q}V_{r,p} + I_{\PM,j,q} + \frac{\widetilde{S}_{\PM,j,q}^{*}}{V_{r,p}^{*}}
	\label{Eq:I}
	\\
	\widetilde{Y}_{\PM,j,q}
	&\coloneq	\frac{V_{j,q}}{V_{r,p}}Y_{\PM,j,q}
	\\
	\widetilde{S}_{\PM,j,q}
	&\coloneq	\frac{V_{r,p}}{V_{j,q}}S_{\PM,j,q}
\end{align}
For convenience, introduce
\begin{align}
	\widetilde{\Y}_{\PM,j}
	&\coloneq \col_{q\in\Set{P}}(\widetilde{Y}_{\PM,j,q})
	\\
	\I_{\PM,j}
	&\coloneq \col_{q\in\Set{P}}(I_{\PM,j,q})
	\\
	\widetilde{\Tensor{S}}_{\PM,j}
	&\coloneq \col_{q\in\Set{P}}(\widetilde{S}_{\PM,j,q})
\end{align}
so that \eqref{Eq:V:1} can be expressed as
%
\begin{align}
	V_{r,p}
	&=	-a_{r,p}V_{r,p} + b_{r,p} + \frac{c_{r,p}}{V_{r,p}^{*}},~
		\begin{array}{l}
			\forall r\in\Set{R},\\
			\forall p\in\Set{P}
		\end{array}
	\label{Eq:V:2}
	\\
	a_{r,p}
	&\coloneq	\sum\limits_{j\in\Set{R}}\row_{p}(\Kron{\Tensor{H}}'_{\mathit{rj}})\widetilde{\Y}_{\PM,j}
	\label{Eq:CQE:a}
	\\
	b_{r,p}
	&\coloneq	\sum\limits_{j\in\Set{R}}\row_{p}(\Kron{\Tensor{H}}'_{\mathit{rj}})\I_{\PM,j}
				+	\widetilde{V}_{\TE,r,p}
	\label{Eq:CQE:b}
	\\
	c_{r,p}
	&\coloneq	\sum\limits_{j\in\Set{R}}\row_{p}(\Kron{\Tensor{H}}'_{\mathit{rj}})\widetilde{\Tensor{S}}_{\PM,j}^{*}
	\label{Eq:CQE:c}
\end{align}


The above-stated equation \eqref{Eq:V:2} can be rearranged to
\begin{equation}
	|V_{r,p}|^{2} - \frac{b_{r,p}}{1+a_{r,p}}V_{r,p}^{*} = \frac{c_{r,p}}{1+a_{r,p}}
\end{equation}
As shown in \cite{J:PSSA:PFS:1986:Kessel}, a complex quadratic equation of this form has a solution if the index $L_{r,p}$, which is defined as
\begin{equation}
	L_{r,p}
	\coloneqq	\left| 1-\frac{b_{r,p}}{1+a_{r,p}}\frac{1}{V_{r,p}} \right|
	=				\left| \frac{c_{r,p}}{1+a_{r,p}} \frac{1}{V_{r,p}^{2}} \right|
	\label{Eq:Index:Local}
\end{equation}
lies in the range
\begin{equation}
	L_{r,p} \leq 1
\end{equation}
In that sense, the indices $L_{r,p}$ are indicators for the solvability of the power-flow equations \eqref{Eq:System:Original:Ohm}--\eqref{Eq:System:Original:Polynomial}.
That is, the power-flow equations are solvable if $L_{r,p}\leqslant1$ $\forall r\in\Set{R}$, $\forall p\in\Set{P}$.
A critical point is reached if one of these local indices equals $1$.
Hence, a global index for static voltage stability is given by
\begin{equation}
	L\coloneq\max\limits_{r\in\Set{R}}\max_{p\in\Set{P}}L_{r,p}
	\label{Eq:Index:Global}
\end{equation}
$L<1$ in the stable region and $L=1$ on the stability boundary.
It is worth noting that, if $\lambda_{r,p}=0$ $\forall r\in\Set{R}$ and $\forall p\in\Set{R}$ (i.e., $\I_{\Set{R}}=\Tensor{0}$), then $L=0$.
If no short-circuit faults occur, which means $V_{r,p}\neq0$ $\forall r\in\Set{R}$ and $\forall p\in\Set{R}$, then the $L_{r,p}$ and $L$ vary continuously in function of the resource parameters.

	
Suppose that the node voltages $V_{r,p}$ and resource parameters $Y_{\PM,r,p}$, $I_{\PM,r,p}$, and $S_{\PM,r,p}$ are known.
Then, the calculation of the \VSI merely requires:
i) a Schur complement for $\Kron{\Tensor{H}}'$ \eqref{Eq:System:Augmented:Hybrid},
ii) multiplications and divisions for $\widetilde{Y}_{\PM,j,q}$ and $\widetilde{S}_{\PM,j,q}$ \eqref{Eq:I},
iii) inner products for $a_{r,p}$, $b_{r,p}$, and $c_{r,p}$ \eqref{Eq:V:2},
iv) additions, divisions, and absolute values for $L_{r,p}$ \eqref{Eq:Index:Local}, and
v) a maximum value for $L$ \eqref{Eq:Index:Global}.
Moreover, the calculation is non-iterative.
Therefore, the proposed \VSI is computationally less intensive than \VSA methods based on \CPF or \NLP[s] (see Sec.~\ref{Sec:Review}).


Finally, observe that the proposed definitions \eqref{Eq:Index:Local} \& \eqref{Eq:Index:Global} are analogous to (20) \& (21) in \cite{J:PSSA:PFS:1986:Kessel}.
In contrast to the original $L$-index \cite{J:PSSA:PFS:1986:Kessel} and the existing extended formulations \cite{J:PSSA:PFS:2005:Hongjie,J:PSSA:PFS:2013:Wang}, \eqref{Eq:Index:Local} \& \eqref{Eq:Index:Global} apply to more generic systems, namely unbalanced polyphase power systems with slack nodes represented by \TE[s] and voltage-dependent power injections represented by \PM[s].
Hypotheses \ref{Hyp:Neutral}--\ref{Hyp:Thevenin} and Theorems~\ref{Thm:Kron}--\ref{Thm:Hybrid} ensure the existence of the compound hybrid parameters needed to compute the \VSI.

\newpage

%% file: Figures/Model_System.tex
\begin{circuitikz}
	\scriptsize
	\def\dx{1.0}
	\def\y{2.5}	
	
	
	\node[draw,minimum width=2cm,minimum height=3cm] (G) {$\Tensor{Y}$};
	
	\coordinate (GAU) at ($(G.west)+(0,0.5*\y)$);
	\coordinate (GAL) at ($(G.west)-(0,0.5*\y)$);
	\coordinate (GBU) at ($(G.east)+(0,0.5*\y)$);
	\coordinate (GBL) at ($(G.east)-(0,0.5*\y)$);
	

	\coordinate (TEU) at ($(GAU)-(0.5*\dx,0)$);
	\coordinate (TEL) at ($(GAL)-(0.5*\dx,0)$);
	\coordinate (TIU) at ($(TEU)-(2*\dx,0)$);
	\coordinate (TIL) at ($(TEL)-(2*\dx,0)$);
	\coordinate (TSU) at ($(TIU)-(\dx,0)$);
	\coordinate (TSL) at ($(TIL)-(\dx,0)$);
		
	\draw (GAU)
		to[short] (TEU)
		to[generic=$\Tensor{Z}_{\TE,s}$,o-] (TIU)
		to[open] (TIL)
		to[short,-o] (TEL)
		to[short] (GAL);
	
	\draw (TIL)
		to[short,o-] (TSL)
		to[voltage source] (TSU)
		to[short,i=$\Tensor{I}_{s}$,-o] (TIU);
	
	
	\coordinate (PEU) at ($(GBU)+(0.5*\dx,0)$);
	\coordinate (PEL) at ($(GBL)+(0.5*\dx,0)$);
	\coordinate (PSU) at ($(PEU)+(1.5*\dx,0)$);
	\coordinate (PSL) at ($(PEL)+(1.5*\dx,0)$);
	
	\draw (GBL)
		to[short,-o] (PEL)
		to[short] (PSL)
		to[controlled current source] (PSU)
		to[short,i=$I_{r,p}$,-o] (PEU)
		to[short] (GBU);
	
	
	\draw (TIL) to[open,v=$\Tensor{V}_{\TE,s}$] (TIU);
	\draw (TEL) to[open,v^=$\Tensor{V}_{s}$] (TEU);
	\draw (PEL) to[open,v=$V_{r,p}$] (PEU);
	
	\draw[decoration={brace,mirror,raise=13pt},decorate] (TIL) -- node[below=18pt] {Augmented Electrical Grid} (PEL);
	\draw[decoration={brace,mirror,raise=12pt},decorate] (TEU) -- node[above=16pt] {Th{\'e}venin Equivalents} (TSU);
	\draw[decoration={brace,mirror,raise=12pt},decorate] (PSU) -- node[above=16pt] {Polynomial Models} (PEU);
	
\end{circuitikz}

%% file: Sections/Validation.tex
\section{Validation}
\label{Sec:Validation}


\subsection{Benchmark System}



\begin{figure}[t]
	\centering
	\input{"Figures/Benchmark_System"}
	\caption{Schematic of the benchmark system.}
	\label{Fig:Benchmark}
\end{figure}
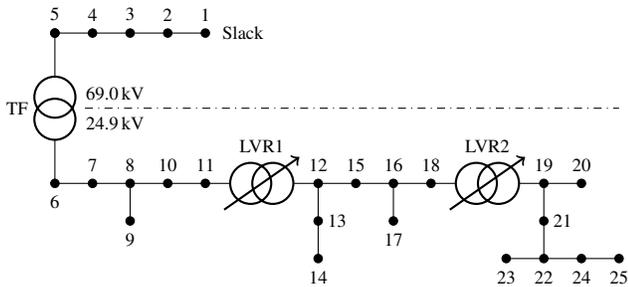

The benchmark system used for the performance evaluation is triphase, and consists of two parts: an upper-level subsystem (nodes 1--5) with nominal voltage 69.0\,kV~phase-to-phase, and a lower-level subsystem (nodes 6--25) with nominal voltage 24.9\,kV~phase-to-phase (see Fig.~\ref{Fig:Benchmark}).
The latter is adapted from the \IEEE 34-node feeder \cite{J:PSM:1991:Kersting}, which contains untransposed overhead lines and \emph{Line Voltage Regulators} (\LVR[s]).
This grid has been chosen for the sake of reproducibility of the results (i.e., because the parameters of this benchmark power system are fully documented and publicly available).



\begin{table}[t]
	\centering
	\caption{Configuration of the Overhead Lines}
	\label{Tab:Lines:Configuration}
	\input{Tables/Lines_Configuration}
\end{table}

\begin{table}[t]
	\centering
	\caption{Sequence Parameters of the Transposed Lines}
	\label{Tab:Lines:Parameters}
	\input{Tables/Lines_Parameters}
\end{table}

\begin{table}[t]
	\centering
	\caption{Configuration of the Transformers}
	\label{Tab:Transformers:Configuration}
	\input{Tables/Transformers_Configuration}
\end{table}

The electrical grid is built of overhead lines (see Tabs.~\ref{Tab:Lines:Configuration}--\ref{Tab:Lines:Parameters}), both transposed and untransposed ones, and transformers (see Tab.~\ref{Tab:Transformers:Configuration}), both regular ones and \LVR[s].
All transformers are wye-connected and effectively grounded both on the primary and secondary side.
Therefore, the sequence impedances are equal.
Here, typical values $R\approx5$E$-3$~p.u. and $X\approx0.1$~p.u. (w.r.t. the base impedance defined by the rated power and the nominal voltage) are used \cite{B:PSP:1985:Roeper}.
The tap ratios of the \LVR[s] are fixed to 1.05 for the sake of simplicity.
If the tap positions are changed (e.g., due to voltage control), one can simply rebuild the compound admittance matrix and compound hybrid matrix, respectively\footnote{Note that transformers (e.g., \LVR[s] or substation transformers with on-load tap changers) are part of the equivalent circuit of the grid.}.



\begin{table}[b]
	\centering
	\caption{Reference Values of the Polynomial Models.}
	\label{Tab:Polynomial:Reference}
	\input{Tables/Resources_Configuration}
\end{table}

\begin{table}[b]
	\centering
	\caption{Normalized Coefficients of the Polynomial Models.}
	\label{Tab:Polynomial:Coefficients}
	\input{Tables/Resources_Parameters}
\end{table}

The slack node is the primary substation (i.e., node~1).
Its \TE consists of a positive-sequence voltage source, which is defined by the rated voltage, and a diagonal compound impedance matrix with equal diagonal entries, which are given by the short-circuit parameters.
The substation is characterized by the short-circuit power $S_{sc}=100$\,MVA and the resistance-to-reactance ratio $R_{sc}/X_{sc}=0.1$.
The resource nodes are in the lower-level subsystem, and host loads and compensators.
Generators are not considered, as static voltage instability due to generation is unlikely in a lossy grid (see Sec.~\ref{Sec:Review:CPF}).
The \PM[s] are specified in Tabs.~\ref{Tab:Polynomial:Reference}--\ref{Tab:Polynomial:Coefficients}.
The load coefficients are taken from \cite{J:PSMI:1988:Price} (i.e., the means of zones 11--16/21--26).
These values are derived from real measurements.
The compensators are \emph{Static Synchronous Compensators} (\STATCOM[s]), which supply constant reactive power \cite{B:PSC:2000:Hingorani} (i.e., $\alpha=\beta=0$, $\gamma=1$).


\subsection{Study Description}

The proposed \VSI is validated by a classical \CPF method.
More precisely, it is verified that the \VSI correctly identifies the loadability limit along the trajectory of the \CPF.

%
Let $\Tensor{\lambda}$ contain the loading factors.
Merging Ohm's law \eqref{Eq:Ohm}, the zero injections \eqref{Eq:Node:Zero}, the \TE[s] \eqref{Eq:Thevenin}, and the \PM[s] \eqref{Eq:Polynomial:P}--\eqref{Eq:Polynomial:Q} yields the power-flow equations as
\begin{equation}
	\Delta\Tensor{S}(\Tensor{V},\Tensor{\lambda}) = \Tensor{0}
	\label{eq:pfe:1}
\end{equation}
where $\Delta\Tensor{S}$ is the mismatch between the nodal injected powers calculated by the grid model and the node models, respectively.
Express $\Delta\Tensor{S}$ in rectangular and $\Tensor{V}$ in polar coordinates:
\begin{align}
	\Delta\Tensor{S}(\Tensor{V},\Tensor{\lambda})
	&\coloneqq	\Delta\Tensor{P}(\Tensor{V},\Tensor{\lambda}) + j\Delta\Tensor{Q}(\Tensor{V},\Tensor{\lambda})
	\\
	\Tensor{V}
	&\coloneqq	\Tensor{E}\angle\Tensor{\theta}
\end{align}
Thus, \eqref{eq:pfe:1} can be restated as a system of real-valued equations in real-valued variables, namely
\begin{equation}
	\left[
	\begin{array}{l}
		\Delta\Tensor{P}(\Tensor{E},\Tensor{\theta},\Tensor{\lambda})	\\
		\Delta\Tensor{Q}(\Tensor{E},\Tensor{\theta},\Tensor{\lambda})
	\end{array}
	\right]
	=	\Tensor{0}
	\label{eq:pfe:2}
\end{equation}
Furthermore, define
\begin{align}
	\Tensor{f}(\Tensor{E},\Tensor{\theta},\Tensor{\lambda})
	&\coloneqq	\left[
						\begin{array}{l}
							\Delta\Tensor{P}(\Tensor{E},\Tensor{\theta},\Tensor{\lambda})	\\
							\Delta\Tensor{Q}(\Tensor{E},\Tensor{\theta},\Tensor{\lambda})
						\end{array}
						\right]
	\\
	\Tensor{x}
	&\coloneqq	\left[
						\begin{array}{l}
							\Tensor{E}	\\
							\Tensor{\theta}
						\end{array}
						\right]
\end{align}
and suppose that $\Tensor{\lambda}$ follows a trajectory parameterized as $\Tensor{\lambda}(\xi)$.
Then, \eqref{eq:pfe:2} can be written compactly as
\begin{equation}
	\Tensor{f}(\Tensor{x},\xi) = \Tensor{0}
\end{equation}


\begin{algorithm}[t]
	\caption{Homotopy continuation method.}
	\label{alg:hcm}
	\input{Algorithms/Homotopy_Continuation_Method}
\end{algorithm}

\begin{algorithm}[t]
	\caption{Newton-Raphson method.}
	\label{alg:nrm}
	\input{Algorithms/Newton_Raphson_Method}
\end{algorithm}

To find the loadability limit $\xi_{\max}$ along the trajectory $\Tensor{\lambda}(\xi)$, one needs to solve the optimization problem
\begin{equation}
	\max\xi~\text{s.t.}~\Tensor{f}(\Tensor{x},\xi)=\Tensor{0}
\end{equation}
It can reasonably be supposed that $\Tensor{f}(\Tensor{x},\xi)$ is continuous \cite{J:PSSA:ML:2009:Avalos}.
Therefore, this maximization problem can be solved using a continuation method.
In this paper, the homotopy continuation method given in Alg.~\ref{alg:hcm}, which is based on \cite{J:PSSA:CPF:1995:Chiang}, is employed.
The continuation step consists of a tangent predictor, which extrapolates guesses $\Tensor{x}^{-}_{k+1}$/$\xi^{-}_{k+1}$ of the next solutions in the continuum, and the Newton-Raphson corrector given in Alg.~\ref{alg:nrm}, which determines the actual values $\Tensor{x}^{+}_{k+1}$/$\xi^{+}_{k+1}$.
%
%
%
$\diff_{\Tensor{x}}$ and $\diff_{\xi}$ are the differential operators\footnote{Observe that, in view of the assumed continuity of the function $\Tensor{f}(\Tensor{x},\xi)$, the derivatives $\diff_{\Tensor{x}}\Tensor{f}(\Tensor{x},\xi)$ and $\diff_{\xi}\Tensor{f}(\Tensor{x},\xi)$ exist.} w.r.t. $\Tensor{x}$ and $\xi$, and $\sigma$ the length of the continuation step.
Following common practice in \VSA, $\Tensor{\lambda}(\xi)$ is chosen as uniform load increase \cite{J:PSSA:CPF:2005:Zhang,J:PSSA:ML:2009:Avalos}.
In other words, $\lambda_{r,p}=\xi$ for the loads and $\lambda_{r,p}=1$ for the compensators.

%
At the loadability limit found by the \CPF method, the \VSI at the critical phase of the critical node must (approximately) equal 1.
Moreover, the loadability limit is verified graphically and numerically as a double-check.
For the graphical analysis, the nose curves of the system and the characteristic curves of the loads are plotted.
These curves are tangent at the critical point.
For the numerical analysis, the singular values of the Jacobian matrix of the power-flow equations are computed.
As the system approaches the critical point, the Jacobian matrix becomes closer to singular.
Thus, at least one singular value tends to zero.


\subsection{Result Discussion}



\begin{figure}[t]
	\centering
	\includegraphics[width=1.0\linewidth]{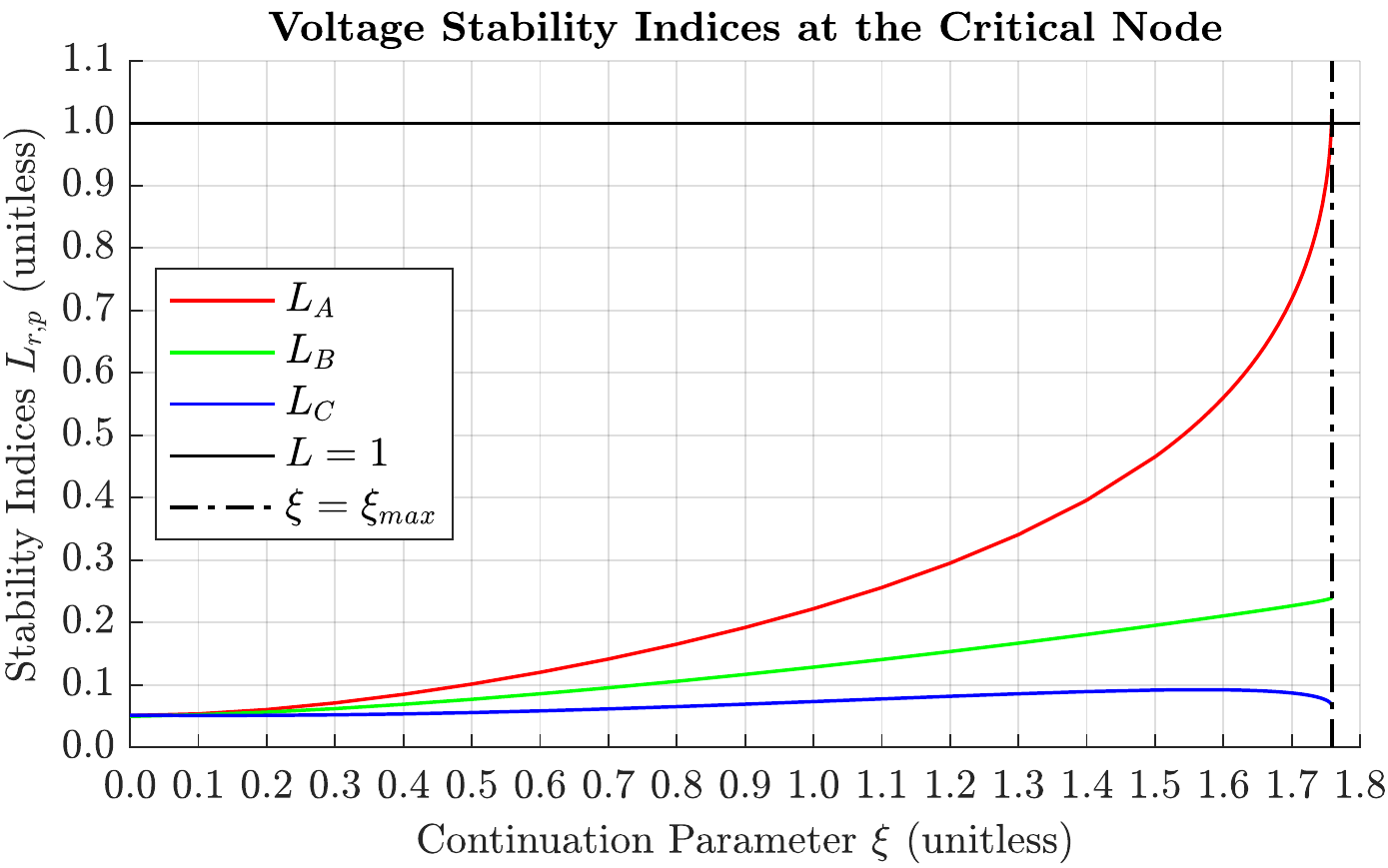}
	\caption{Evolution of the \VSI[s] at the critical node (i.e., node~25).}
	\label{Fig:Result:Index}
\end{figure}

The loadability limit lies at $\xi_{\max}\approx1.759$.
The maximum value of the \VSI occurs in phase~$A$ of node~25: $L_{25,A}=1.017$.
This point in the grid has the highest load (see Tab.~\ref{Tab:Polynomial:Reference}), and is furthest away from the slack (see Fig.~\ref{Fig:Benchmark}).
The evolution of the \VSI[s] at this node is shown in Fig.~\ref{Fig:Result:Index}.
Clearly, only $L_{25,A}$ tends to $1$ as $\xi$ increases, whereas $L_{25,B}$ and $L_{25,C}$ remain much lower.
This behavior is consistent with what has been observed for original $L$-index in \cite{J:PSSA:PFS:1986:Kessel}.
The \VSI[s] in the other nodes of the system behave similarly.
That is, the indices in phase~$A$ are higher than those in phases~$B$ and $C$, and all of them are lower than those in node~25.



\begin{figure}[t]
	\centering
	\includegraphics[width=1.0\linewidth]{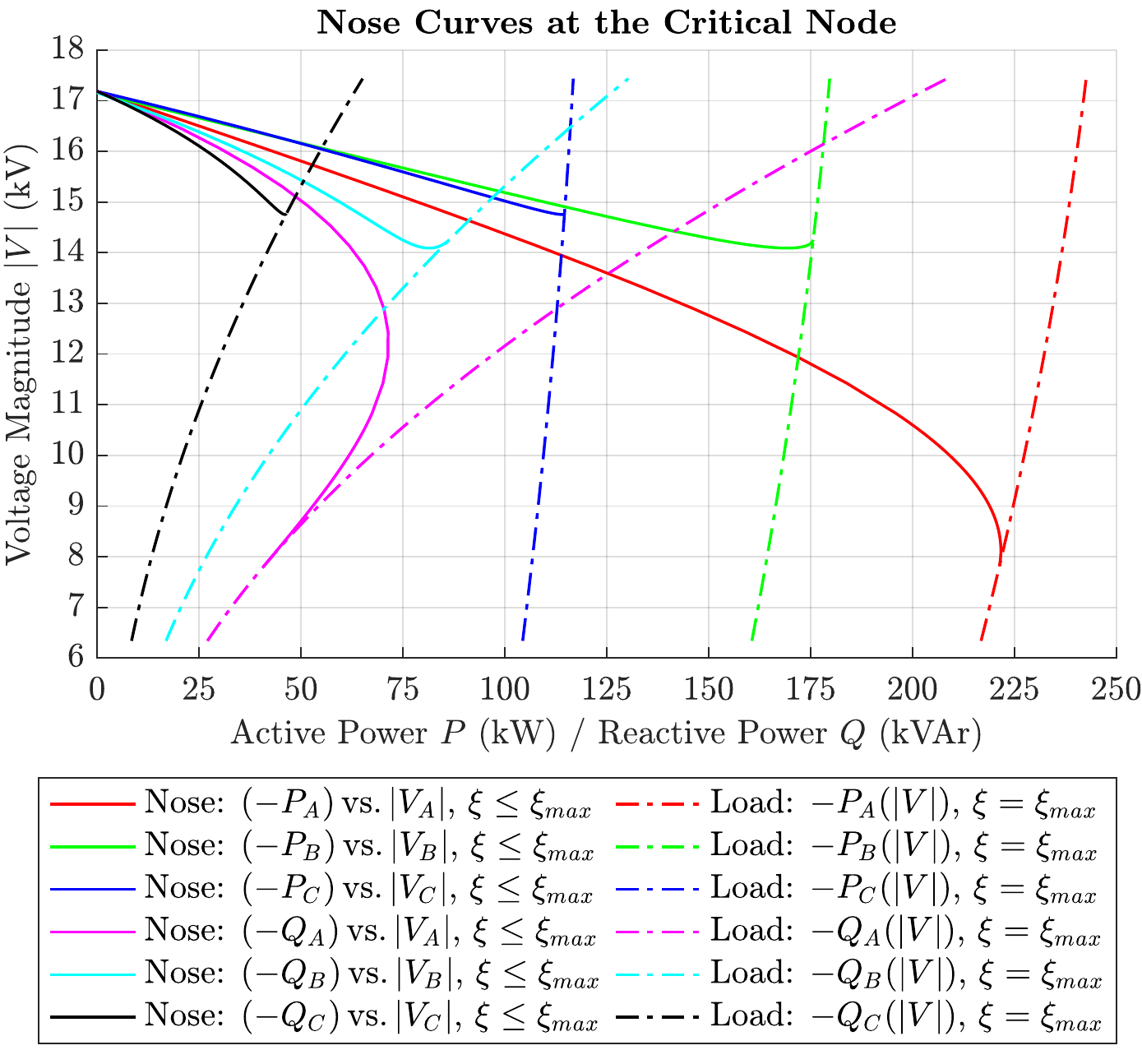}
	\caption{Nose curves and load characteristics at the critical node (i.e., node~25).}
	\label{Fig:Result:Nose}
\end{figure}

The nose curves of the system (for $\xi\leqslant\xi_{\max}$) and the characteristic curves of the load (for $\xi=\xi_{\max}$) at the critical node are depicted in Fig.~\ref{Fig:Result:Nose}.
Evidently, these curves are tangent for phase $A$.
So, the graphical analysis confirms the results of the \CPF method and the \VSI.
Incidentally, it is worthwhile mentioning that the nose curves of phase $A$ are bent downward (i.e., towards lower voltage), whereas those of phases $B$ and $C$ start bending upwards as $\xi$ approaches $\xi_{\max}$.
In phase $B$, the change of curvature is clearly visible.
This behavior is in accordance with the \CPF analysis of unbalanced triphase systems in \cite{J:PSSA:CPF:2005:Zhang,J:PSSA:CPF:2014:Sheng}.



\begin{figure}[t]
	\centering
	\includegraphics[width=1.0\linewidth]{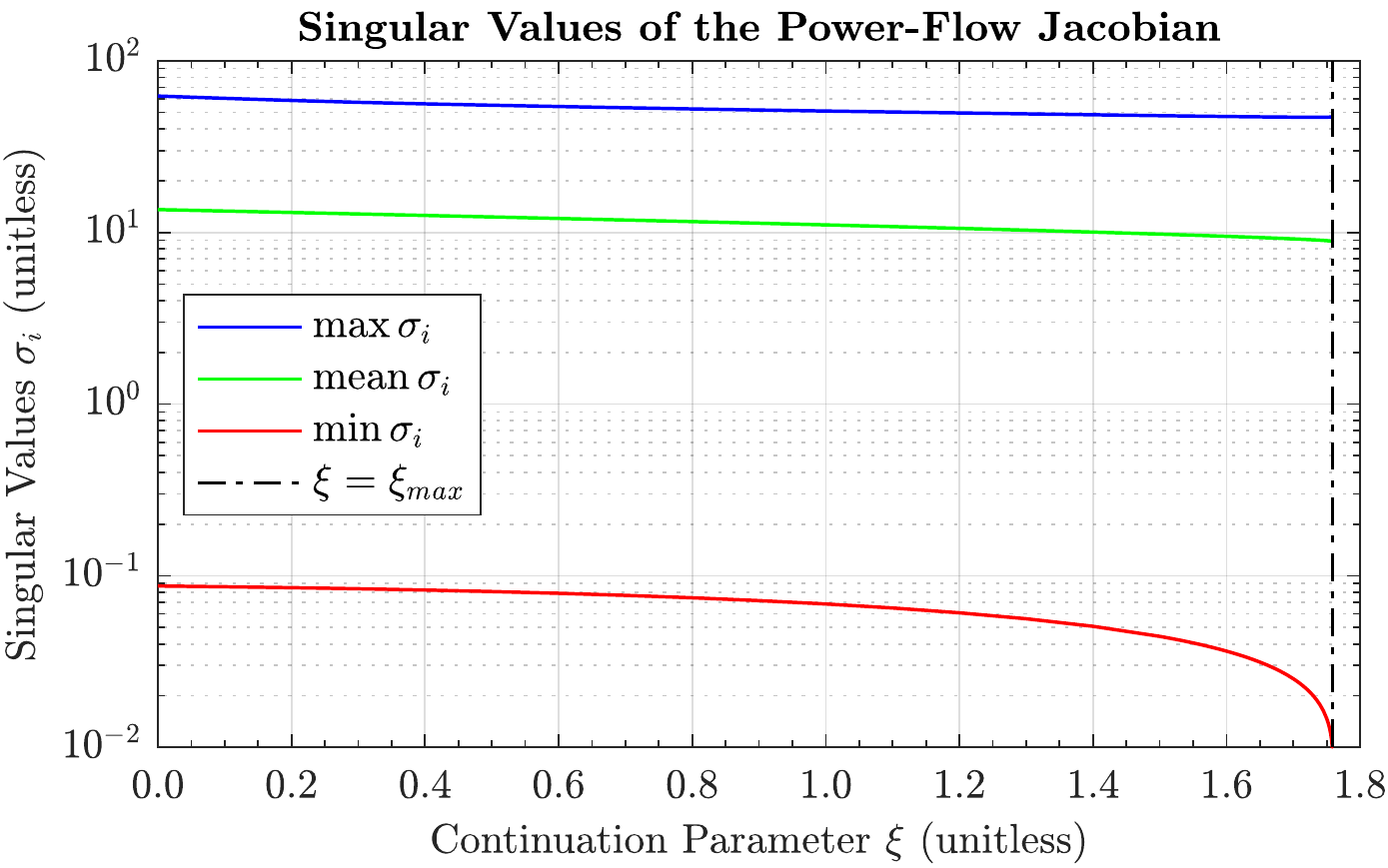}
	\caption{Evolution of the singular values of the Power-Flow Jacobian.}
	\label{Fig:Result:Jacobian}
\end{figure}

The evolution of the maximum, minimum, and mean of the singular values of the power-flow Jacobian is shown in Fig.~\ref{Fig:Result:Jacobian}.
Obviously, the maximum and mean value remain almost constant over the entire range of $\xi$, whereas the minimum value plummets as $\xi_{\max}$ is approached.
This means that the power-flow Jacobian is virtually singular at $\xi_{\max}$.
This is also in agreement with the results obtained using the \CPF method and the \VSI.
So, the \VSI detects the instability correctly.



Finally, there are some comments to be made regarding the practicality of the obtained results.
It can be seen in Fig.~\ref{Fig:Result:Nose} that the voltage in phase~$A$ of node~25 is low: roughly 8\,kV, or around 55\% of the nominal voltage.
This value is outside the range desired for regular operation.
According to Tab.~\ref{Tab:Nodes:Voltages}, low voltages only occur in phase~$A$ of the load nodes, where the load is higher (see Tab.~\ref{Tab:Polynomial:Reference}).
In phases~$B$ and $C$, in contrast, the voltages are close to the nominal value.
Moreover, according to Tab.~\ref{Tab:Lines:Currents}, the thermal line ratings are respected with ample margin throughout the system.
In view of the obtained results, it can be concluded that the identified loadability limit is of practical interest.
Finally, it is worth noting that static voltage instability may well occur at close-to-nominal voltage in power distribution systems \cite{J:PSSA:PFS:1998:Prada} (i.e., depending on the grid and load).
This confirms the need for an accurate assessment of the static voltage stability.
	
\begin{table}[b]
	\centering
	\caption{Voltage Magnitudes at the Load Nodes at $\xi=\xi_{\max}$.}
	\input{Tables/Nodes_Voltages}
\end{table}

\begin{table}[b]
	\centering
	\caption{Conductor Currents of Selected Lines at $\xi=\xi_{\max}$.}
	\label{Tab:Lines:Currents}
	\input{Tables/Lines_Currents}
\end{table}

\newpage

%% file: Figures/Benchmark_System.tex
\begin{circuitikz}
	\scriptsize
	
	\def\dx{0.5}
	\def\dy{0.5}
	
	
	\coordinate (N1) at (0,0);
	\coordinate (N2) at ($(N1)-(\dx,0)$);
	\coordinate (N3) at ($(N2)-(\dx,0)$);
	\coordinate (N4) at ($(N3)-(\dx,0)$);
	\coordinate (N5) at ($(N4)-(\dx,0)$);
	\coordinate (N6) at ($(N5)-(0,4.0*\dy)$);
	\coordinate (N7) at ($(N6)+(\dx,0)$);
	\coordinate (N8) at ($(N7)+(\dx,0)$);
	\coordinate (N9) at ($(N8)-(0,\dy)$);
	\coordinate (N10) at ($(N8)+(\dx,0)$);
	\coordinate (N11) at ($(N10)+(\dx,0)$);
	\coordinate (N12) at ($(N11)+(3.0*\dx,0)$);
	\coordinate (N13) at ($(N12)-(0,\dy)$);
	\coordinate (N14) at ($(N13)-(0,\dy)$);
	\coordinate (N15) at ($(N12)+(\dx,0)$);
	\coordinate (N16) at ($(N15)+(\dx,0)$);
	\coordinate (N17) at ($(N16)-(0,\dy)$);
	\coordinate (N18) at ($(N16)+(\dx,0)$);
	\coordinate (N19) at ($(N18)+(3.0*\dx,0)$);
	\coordinate (N20) at ($(N19)+(\dx,0)$);
	\coordinate (N21) at ($(N19)-(0,\dy)$);
	\coordinate (N22) at ($(N21)-(0,\dy)$);
	\coordinate (N23) at ($(N22)-(\dx,0)$);
	\coordinate (N24) at ($(N22)+(\dx,0)$);
	\coordinate (N25) at ($(N24)+(\dx,0)$);
	
	\coordinate (TF) at ($0.5*(N5)+0.5*(N6)$);
	\coordinate (LVR1) at ($0.5*(N11)+0.5*(N12)$);
	\coordinate (LVR2) at ($0.5*(N18)+0.5*(N19)$);	
	
	
	\draw (N1)
		to[short,*-] (N2)
		to[short,*-] (N3)
		to[short,*-] (N4)
		to[short,*-*] (N5)
		to[voosource,*-](N6)
		to[short,*-] (N7)
		to[short,*-] (N8)
		to[short,*-] (N10)
		to[short,*-] (N11)
		to[voosource,*-] (N12)
		to[short,*-] (N15)
		to[short,*-] (N16)
		to[short,*-] (N18)
		to[voosource,*-] (N19)
		to[short,*-] (N21)
		to[short,*-] (N22)
		to[short,*-] (N24)
		to[short,*-*] (N25);
		
	\draw (N8) to[short,-*] (N9);
	\draw (N12)
		to[short,-*] (N13)
		to[short,-*] (N14);
	\draw (N16) to[short,-*] (N17);
	\draw (N19) to[short,-*] (N20);
	\draw (N22) to[short,-*] (N23);
	
	
	\node at ($(N1)+(0,0.5*\dy)$) {1};
	\node at ($(N2)+(0,0.5*\dy)$) {2};
	\node at ($(N3)+(0,0.5*\dy)$) {3};
	\node at ($(N4)+(0,0.5*\dy)$) {4};
	\node at ($(N5)+(0,0.5*\dy)$) {5};
	\node at ($(N6)-(0,0.5*\dy)$) {6};
	\node at ($(N7)+(0,0.5*\dy)$) {7};
	\node at ($(N8)+(0,0.5*\dy)$) {8};
	\node at ($(N9)-(0,0.5*\dy)$) {9};
	\node at ($(N10)+(0,0.5*\dy)$) {10};
	\node at ($(N11)+(0,0.5*\dy)$) {11};
	\node at ($(N12)+(0,0.5*\dy)$) {12};
	\node at ($(N13)+(0.5*\dx,0)$) {13};
	\node at ($(N14)-(0,0.5*\dy)$) {14};
	\node at ($(N15)+(0,0.5*\dy)$) {15};
	\node at ($(N16)+(0,0.5*\dy)$) {16};
	\node at ($(N17)-(0,0.5*\dy)$) {17};
	\node at ($(N18)+(0,0.5*\dy)$) {18};
	\node at ($(N19)+(0,0.5*\dy)$) {19};
	\node at ($(N20)+(0,0.5*\dy)$) {20};
	\node at ($(N21)+(0.5*\dx,0)$) {21};
	\node at ($(N22)-(0,0.5*\dy)$) {22};
	\node at ($(N23)-(0,0.5*\dy)$) {23};
	\node at ($(N24)-(0,0.5*\dy)$) {24};
	\node at ($(N25)-(0,0.5*\dy)$) {25};
	
	\node at ($(N1)+(\dx,0)$) {Slack};
	\node at ($(TF)-(\dx,0)$) {TF};
	\node at ($(TF)+(1.6*\dx,0.4*\dy)$) {69.0\,kV};
	\node at ($(TF)+(1.6*\dx,-0.4*\dy)$) {24.9\,kV};
	
	\draw [->,thick] ($(LVR1)-(\dx,0.7*\dy)$) to ($(LVR1)+(\dx,0.7*\dy)$);
	\node at ($(LVR1)+(0,\dy)$) {\LVR[1]};
	
	\draw [->,thick] ($(LVR2)-(\dx,0.7*\dy)$) to ($(LVR2)+(\dx,0.7*\dy)$);
	\node at ($(LVR2)+(0,\dy)$) {\LVR[2]};
	
	\draw[dashdotted] ($(N7)+(-0.2*\dx,2.0*\dy)$) to ($(N20)+(\dx,2.0*\dy)$);
\end{circuitikz}

%% file: Tables/Lines_Configuration.tex
{
\renewcommand{\arraystretch}{1.2}
\begin{tabular}{ccccc}
	\hline
	Line							&Length					&Parameters									&Transposed?\\
	(Nodes)						&(km)						&													&(Yes/No)\\
	\hline
	1$-$2						&25.000					&Tab.~\ref{Tab:Lines:Parameters}	&Yes\\
	2$-$3						&25.000					&Tab.~\ref{Tab:Lines:Parameters}	&Yes\\
	3$-$4						&25.000					&Tab.~\ref{Tab:Lines:Parameters}	&Yes\\
	4$-$5						&25.000					&Tab.~\ref{Tab:Lines:Parameters}	&Yes\\
	6$-$7						&\hphantom{0}1.314	&\IEEE-300									&No\\
	7$-$8						&\hphantom{0}9.851	&\IEEE-300									&No\\
	8$-$9						&\hphantom{0}1.769	&\IEEE-300									&No\\
	\hphantom{0}8$-$10	&11.430						&\IEEE-300									&No\\
	10$-$11						&\hphantom{0}9.062	&\IEEE-300									&No\\
	12$-$13						&15.197					&\IEEE-301									&No\\
	13$-$14						&\hphantom{0}4.188	&\IEEE-301									&No\\
	12$-$15						&\hphantom{0}3.112	&\IEEE-301									&No\\
	15$-$16						&\hphantom{0}6.645	&\IEEE-301									&No\\
	16$-$17						&\hphantom{0}7.111	&\IEEE-301									&No\\
	16$-$18						&11.226						&\IEEE-301									&No\\
	19$-$20						&\hphantom{0}3.219	&\IEEE-301									&No\\
	19$-$21						&\hphantom{0}1.494	&\IEEE-301									&No\\
	21$-$22						&\hphantom{0}1.777	&\IEEE-301									&No\\
	22$-$23						&\hphantom{0}1.768	&\IEEE-301									&No\\
	22$-$24						&\hphantom{0}1.433	&\IEEE-301									&No\\
	24$-$25						&\hphantom{0}1.567	&\IEEE-301									&No\\
	\hline
\end{tabular}
}

%% file: Tables/Lines_Parameters.tex
{
\renewcommand{\arraystretch}{1.3}
\begin{tabular}{cccc}
	\hline
	Sequence						&$R'$~($\Omega$/km)	&$X'$~($\Omega$/km)	&$B'$~($\mu$S/km)
	\\
	\hline
	Positive\,$+$\,Negative	&0.071							&0.379							&3.038\\
	Homopolar						&0.202							&0.884							&1.740
	\\
	\hline
\end{tabular}
}

%% file: Tables/Transformers_Configuration.tex
{
\renewcommand{\arraystretch}{1.2}
\begin{tabular}{ccccc}
	\hline
	Name		&Link				&Rated~Power		&Nominal~Voltage\\
					&(Nodes, I$-$II)	&(MVA)					&(kV,~phase-to-phase)\\
	\hline
	TF			&5$-$6				&12.0					&69.0\,(I),~24.9\,(II)\\
	\LVR[1]		&11$-$12			&\hphantom{0}9.0	&24.9\,(I$+$II)\\
	\LVR[2]		&18$-$19			&\hphantom{0}9.0	&24.9\,(I$+$II)\\
	\hline
\end{tabular}
}

%% file: Tables/Resources_Configuration.tex
{
\renewcommand{\arraystretch}{1.2}
\begin{tabular}{ccccc}
	\hline
		Node	
	&	$V_{0}$	&$P_{0,A}$,~$P_{0,B}$,~$P_{0,C}$
	&	$Q_{0,A}$,~$Q_{0,B}$,~$Q_{0,C}$
	&	Type
	\\
	&	(kV)
	&	(kW)
	&	(kVAR)\\			
	\hline
		9
	&	14.4
	&	\hphantom{0}$-$60,~\hphantom{0}$-$50,~\hphantom{0}$-$40
	&	\hphantom{0}$-$30,~\hphantom{0}$-$25,~\hphantom{0}$-$20
	&	Load
	\\
		14	
	&	14.4
	&	\hphantom{0}$-$75,~\hphantom{0}$-$60,~\hphantom{0}$-$45
	&	\hphantom{0}$-$40,~\hphantom{0}$-$30,~\hphantom{0}$-$21
	&	Load
	\\
		17	
	&	14.4
	&	\hphantom{0}$-$90,~\hphantom{0}$-$70,~\hphantom{0}$-$50
	&	\hphantom{0}$-$50,~\hphantom{0}$-$35,~\hphantom{0}$-$22
	&	Load
	\\
		20
	&	14.4
	&	$-$105,~\hphantom{0}$-$80,~\hphantom{0}$-$55
	&	\hphantom{0}$-$60,~\hphantom{0}$-$40,~$-\hphantom{0}$23
	&	Load
	\\
		23
	&	14.4
	&	$-$120,~\hphantom{0}$-$90,~\hphantom{0}$-$60
	&	\hphantom{0}$-$70,~\hphantom{0}$-$45,~\hphantom{0}$-$24
	&	Load
	\\
		25
	&	14.4
	&	$-$135,~$-$100,~\hphantom{0}$-$65
	&	\hphantom{0}$-$80,~\hphantom{0}$-$50,~\hphantom{0}$-$25
	&	Load
	\\
		12
	&	14.4
	&	\hphantom{00$-$}0,~\hphantom{00$-$}0,~\hphantom{00$-$}0
	&	\hphantom{$-$}100,~\hphantom{$-$}100,~\hphantom{$-$}100
	&	Comp.
	\\
		19
	&	14.4
	&	\hphantom{00$-$}0,~\hphantom{00$-$}0,~\hphantom{00$-$}0
	&	\hphantom{$-$}100,~\hphantom{$-$}100,~\hphantom{$-$}100
	&	Comp.
	\\
	\hline
\end{tabular}
}

%% file: Tables/Resources_Parameters.tex
{
\renewcommand{\arraystretch}{1.2}
\begin{tabular}{ccc}
	\hline
		Type
	&	$\alpha_{\Re}$,~$\beta_{\Re}$,~$\gamma_{\Re}$
	&	$\alpha_{\Im}$,~$\beta_{\Im}$,~$\gamma_{\Im}$
	\\
	\hline
		Load					
	&	$-$0.067,~\hphantom{$-$}0.251,~\hphantom{$-$}0.816
	&	\hphantom{$-$}1.064,~$-$0.088,~\hphantom{$-$}0.025
	\\
		Comp.
	&	\hphantom{$-$}0.000,~\hphantom{$-$}0.000,~\hphantom{$-$}0.000
	&	\hphantom{$-$}0.000,~\hphantom{$-$}0.000,~\hphantom{$-$}1.000
	\\
	\hline
\end{tabular}
}

%% file: Algorithms/Homotopy_Continuation_Method.tex
\begin{algorithmic}
	\Procedure{HCM}{$\Tensor{f}(\Tensor{x},\xi)$, $\Tensor{x}_{0}$, $\xi_{0}$}
		\State{\# Solve $\max\xi~\text{s.t.}~\Tensor{f}(\Tensor{x},\xi)=\Tensor{0}$, starting at $\Tensor{x}_{0},\xi_{0}$.}
		\For{k$\geqslant0$}
		\State{\# Predictor (tangent method)}
		\State{$d\Tensor{x} \gets \text{solve}\left(\diff_{\Tensor{x}}\Tensor{f}(\Tensor{x}_{k},\xi_{k})d\Tensor{x}=-\diff_{\xi}\Tensor{f}(\Tensor{x}_{k},\xi_{k}),d\Tensor{x}\right)$}
		\State{%
			$
			\left[
			\begin{array}{l}
				\Tensor{x}^{-}_{k+1}\\
				\xi^{-}_{k+1}
			\end{array}
			\right]
			=		\left[
					\begin{array}{l}
						\Tensor{x}_{k}\\
						\xi_{k}
					\end{array}
					\right]
				+	\sigma\left(\frac{1}{\sqrt{\Norm{d\Tensor{x}}^{2}+1}}
					\left[
					\begin{array}{c}
						d\Tensor{x}\\
						1
					\end{array}
					\right]\right)
			$%
		}
		\vspace{3pt}
		\State{\# Corrector (Newton-Raphson method)}
		\vspace{1pt}
		\State{%
			$
			\Tensor{g}([\Tensor{x};\xi])
			\coloneqq	\left[\hspace{-3pt}
							\begin{array}{l}
								\Tensor{f}(\Tensor{x},\xi)\\
								\Norm{\Tensor{x}-\Tensor{x}_{k}}^{2} + (\xi-\xi_{k})^{2} - \sigma^{2}
							\end{array}
							\hspace{-3pt}\right]
			$%
		}
		\State{$[\Tensor{x}^{+}_{k+1}; \xi^{+}_{k+1}] \gets \text{NRM}\left(\Tensor{g}([\Tensor{x};\xi]), [\Tensor{x}^{-}_{k+1}; \xi^{-}_{k+1}]\right)$}
		\State{%
			$
			\left[
			\begin{array}{l}
				\Tensor{x}_{k+1}\\
				\xi_{k+1}
			\end{array}
			\right]
			\gets	\left[
					\begin{array}{l}
						\Tensor{x}^{+}_{k+1}\\
						\xi^{+}_{k+1}
					\end{array}
					\right]
			$%
		}
		\vspace{2pt}
		\If{$\sign(\xi_{k+1}-\xi_{k})\leqslant0$}\Comment{$\max\xi$ found.}
			\State{\textbf{break}}
		\EndIf
		\EndFor
		\State{\textbf{return} $\{\Tensor{x}_{k}, \xi_{k}\}$}\Comment{Continuum of solutions $\{\Tensor{x}_{k},\xi_{k}\}$.}
	\EndProcedure
\end{algorithmic}

%% file: Algorithms/Newton_Raphson_Method.tex
\begin{algorithmic}
	\Procedure{NRM}{$\Tensor{g}(\Tensor{x})$, $\Tensor{x}_{0}$}
		\State{\# NRM solves $\Tensor{g}(\Tensor{x})=\Tensor{0}$, with initial guess $\Tensor{x}_{0}$.}
		\For{$i\geqslant0$}
			\State{$\Delta\Tensor{g} \gets \Tensor{g}(\Tensor{x}_{i})$}
			\If{$\Norm{\Delta\Tensor{g}}\leqslant\varepsilon$}\Comment{Convergence.}
				\State{\textbf{break}}
			\Else\Comment{Correction step.}
				\State{$\Tensor{J} \gets \diff_{\Tensor{x}}\Tensor{g}(\Tensor{x}_{i})$}
				\State{$\Delta\Tensor{x} \gets \text{solve}\left(\Tensor{J}\Delta\Tensor{x}=\Delta\Tensor{g},\Delta\Tensor{x}\right)$}
				\State{$\Tensor{x}_{i+1} \gets \Tensor{x}_{i} - \Delta\Tensor{x}$}
			\EndIf
		\EndFor
		\State{\textbf{return} $\Tensor{x}_{i}$}\Comment{Solution $\Tensor{x}_{i}$.}
	\EndProcedure
\end{algorithmic}

%% file: Tables/Nodes_Voltages.tex
{
\renewcommand{\arraystretch}{1.2}
\label{Tab:Nodes:Voltages}
\begin{tabular}{ccccc}
	\hline
	Node		&$V_{A}$\,(kV)		&$V_{B}$\,(kV)	&$V_{C}$\,(kV)	&$V_{\text{nominal}}$\,(kV)\\
	\hline
	9			&12.1					&14.1				&14.4				&14.4\\
	14			&\hphantom{0}9.9	&14.1				&14.5				&14.4\\
	17			&\hphantom{0}8.8	&13.9				&14.3				&14.4\\
	20			&\hphantom{0}8.1	&14.3				&14.8				&14.4\\
	23			&\hphantom{0}7.9	&14.3				&14.8				&14.4\\
	25			&\hphantom{9}7.8	&14.3				&14.8				&14.4\\
	\hline
\end{tabular}
}

%% file: Tables/Lines_Currents.tex
{
\renewcommand{\arraystretch}{1.2}
\begin{tabular}{cccccc}
	\hline
	Line							&$I_{A}$\,(A)				&$I_{B}$\,(A)	&$I_{C}$\,(A)			&$I_{\text{rated}}$\,(A)\\
	\hline
	1$-$2						&\hphantom{0}40.8		&21.1			&18.4					&300\\
	5$-$6						&120.6						&60.8			&40.9					&230\\
	\hphantom{0}8$-$10	&111.9						&54.1			&36.1					&230\\
	12$-$15						&\hphantom{0}95.3		&45.5			&29.0					&180\\
	16$-$18						&\hphantom{0}78.3		&36.1			&22.7					&180\\
	19$-$21						&\hphantom{0}54.2		&26.0			&16.0					&180\\
	22$-$24						&\hphantom{0}28.8		&13.7			&\hphantom{0}8.4	&180\\
	\hline
\end{tabular}
}

%% file: Sections/Conclusions.tex
\section{Conclusion}
\label{Sec:Conclusion}

This paper developed a \VSI which is suitable for unbalanced polyphase power systems.
To this end, a system model consisting of polyphase two-port equivalent circuits as well \TE[s] and \PM[s] was formulated.
Using this system model, the power-flow equations were approximated by a system of complex quadratic equations, whose coefficients are calculated from the compound hybrid matrix of the grid and the parameters of the \TE[s] and \PM[s].
The \VSI was derived from the conditions for the solvability of the aforementioned quadratic equations.
In this context, it was illlustrated that the computational burden for the calculation of the \VSI is low.
Finally, the \VSI was validated using a benchmark system based on the \IEEE 34-node feeder.
For this validation, the nose curves of the system and the singular values of the power-flow Jacobian were used.

%% file: Sections/Biographies.tex
\newpage


\begin{IEEEbiography}[{\includegraphics[width=1in,height=1.25in,keepaspectratio]{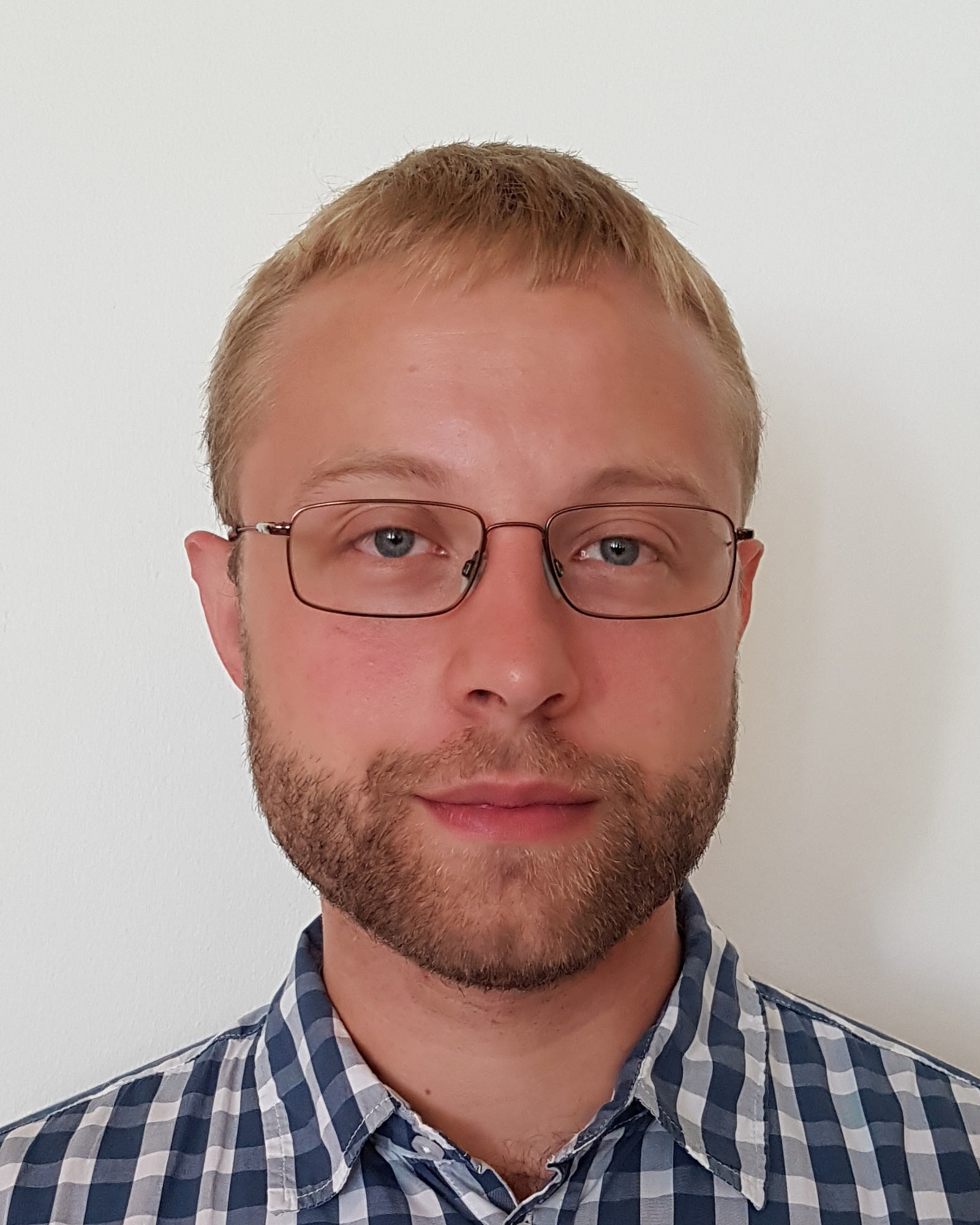}}]{Andreas Martin Kettner}
	(M'15) received the M.Sc. degree in electrical engineering and information technology from the Swiss Federal Institute of Technology of Z{\"u}rich (ETHZ), Z{\"u}rich, Switzerland, in 2014 and the Ph.D. degree in electrical engineering from the Swiss Federal Institute of Technology of Lausanne (EPFL), Lausanne, Switzerland, in 2019.\\
	During 2014, he worked as a Development Engineer at Supercomputing Systems AG, Z{\"u}rich.
	Since 2015, he has been with the Distributed Electrical Systems Laboratory (DESL) at the Swiss Federal Institute of Technology of Lausanne (EPFL), Lausanne, Switzerland, where he is currently a Postdoctoral Researcher.\\
	His research is focused on real-time monitoring and control of power systems, with particular reference to state estimation and voltage stability assessment.
\end{IEEEbiography}



\begin{IEEEbiography}[{\includegraphics[width=1in,height=1.25in,keepaspectratio]{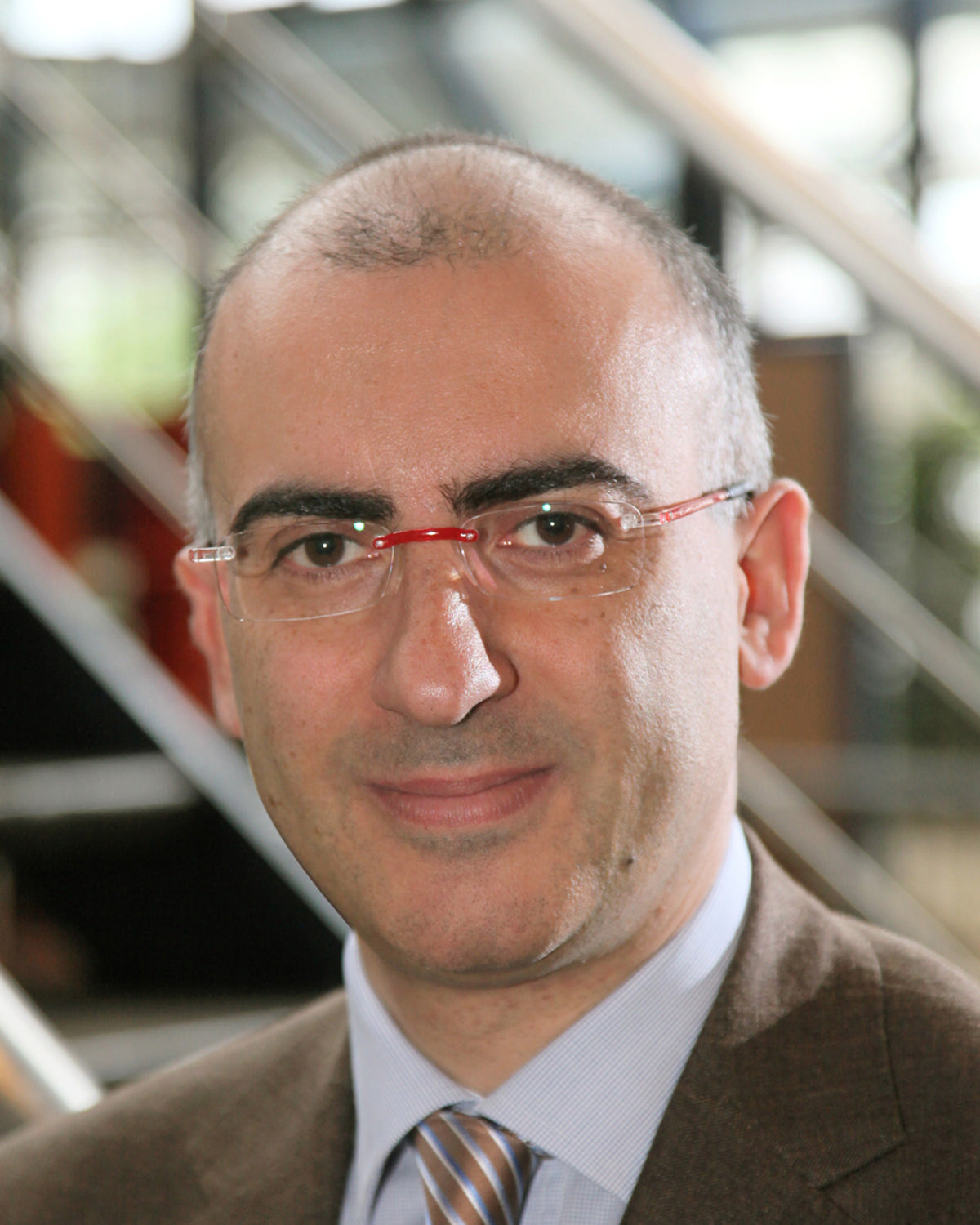}}]{Mario Paolone}
	(M'07-SM'10) received the M.Sc. (with Hons.) and Ph.D. degrees in electrical engineering from the University of Bologna, Italy, in 1998 and 2002, respectively.\\
	In 2005, he was appointed Assistant Professor in power systems with the University of Bologna, where he was with the Power Systems Laboratory until 2011.
	In 2010, he received the Associate Professor eligibility from the Polytechnic of Milan, Italy.
	Since 2011, he is with the Swiss Federal Institute of Technology of Lausanne (EPFL), Lausanne, Switzerland, where he is currently Full Professor, Chair of the Distributed Electrical Systems Laboratory (DESL), Head of SCCER--FURIES (Swiss Competence Center for Energy Research, Future Swiss Electrical Infrastructure), and Chair of the EPFL Energy Centre Directorate.\\
	He has authored or co-authored over 300 scientific papers published in mainstream journals and international conferences in the area of energy and power systems.
	His research interests include power systems with particular reference to real-time monitoring and operation and power system protections, dynamics, and transients.
	He is the Editor-in-Chief of the Elsevier journal Sustainable Energy, Grids and Networks. 
\end{IEEEbiography}

\vfill